\documentclass{article}

\usepackage{arxiv}

\usepackage[utf8]{inputenc} 
\usepackage[T1]{fontenc}    
\usepackage{hyperref}       
\usepackage{url}            
\usepackage{booktabs}       
\usepackage{amsfonts}       
\usepackage{nicefrac}       
\usepackage{microtype}      
\usepackage{lipsum}
\usepackage{graphicx}
\usepackage{enumitem}
\usepackage{float}
\graphicspath{ {./images/} }
\usepackage{amsmath}
\usepackage{multirow}

\title{Considering parallel tempering and comparing post-treatment procedures in Bayesian Profile Regression Models for a survival outcome and correlated exposures}

\author{
 Julie Fendler \\
 Autorité de Sûreté Nucléaire et de Radioprotection (ASNR)\\
 PSE-SANTE/SESANE/LEPID\\
 Fontenay-aux-Roses, France\\
  \texttt{julie.fendler@mrc-bsu.cam.ac.uk} \\
   \And
  Chantal Guihenneuc \footnote[1]{This authors contributed equally} \\
  Université Paris Cité \\
  CNRS, Inserm, P-MIM, BioSTM F-75006\\
  Paris, France\\
  \texttt{chantal.guihenneuc-jouyaux@u-paris.fr} \\
  \And
 Sophie Ancelet \footnote[1]{This authors contributed equally} \\
  Autorité de Sûreté Nucléaire et de Radioprotection (ASNR)\\
 PSE-SANTE/SESANE/LEPID\\
 Fontenay-aux-Roses, France\\
  \texttt{sophie.ancelet@asnr.fr} \\
}

\begin{document}
\maketitle
\begin{abstract}
Bayesian profile regression mixture models (BPRM) allow to assess a health risk in a multi-exposed population. These mixture models cluster individuals according to their exposure profile and their health risk. However, their results, based on Monte-Carlo Markov Chain (MCMC) algorithms, turned out to be unstable in different application cases. We suppose two reasons for this instability. The MCMC algorithm can be trapped in local modes of the posterior distribution and the choice of post-treatment procedures used on the output of the MCMC algorithm leads to different clustering structures. In this work, we propose improvements of the MCMC algorithms proposed in previous works in order to avoid the local modes of the posterior distribution while reducing the computation time. We also carry out a simulation study to compare the performances of the MCMC algorithms and different post-processing in order to provide guidelines on their use. An application in radiation epidemiology is considered.
\end{abstract}


\begingroup
  \renewcommand\thefootnote{}
  \footnotetext{* This authors contributed equally.}
  \addtocounter{footnote}{-1}
\endgroup

\section{Introduction}

In epidemiology, a common problem is to identify subgroups of individuals in a given population that have similar characteristics, for example, environmental exposure conditions or gene expression\cite{HwangLee2005, Ahlqvist2018, Harmouche-Karaki2019, Guillien2021b, Guillien2022, Wade2023}. In statistics, the problem of partitioning individuals into different groups (also called clusters), such that individuals within the same group are more similar for given characteristics than to those in other groups, is referred to as clustering. In addition to clustering individuals according to their observed characteristics, called covariates, another question of interest is whether individuals assigned to the same group also tend to have a more similar level of risk, for one disease of interest, than those in other groups and to identify clusters with higher or lower risks. In radiation epidemiology, for example, we are interested in identifying subgroups of uranium miners sharing similar risk of death by lung cancer and similar professional exposure characteristics to ionising radiation \cite{Rage2015,Belloni2020}. \newline 

To address the above issues, the standard approach is to cluster individuals in a first step and to estimate the health risk associated to each identified cluster in a second step \cite{Majcherek2021, Ma2023, Chuang2024}. However, in health-related applications, clusters are often not well separated and the incertitude on the clustering is often high. An important limitation of the standard approach is that this incertitude is not taken into account when estimating the health risk associated to each cluster. Furthermore, the information contained in the health-related outcome (or response) is not used as an input variable in the clustering algorithm when it could guide the inference algorithm towards a relevant clustering structure. Bayesian Profile Regression Mixture models (BPRM), introduced by Molitor \textit{et al.} (2010)\cite{Molitor2010},  overcome both of these limitations. Inferring these probabilistic models allows both the identification of groups of individuals with similar exposure characteristics (also called profiles) and similar health risks as well as estimation of the associated risk for each group in a unique step. Moreover, all uncertainty, including uncertainty
associated with the clustering structure, is reflected in the estimation of risk parameters. BPRM are increasingly applied in environmental epidemiology \cite{Molitor2011, Hastie2013, Pirani2015, Coker2016, Coker2018, Lavigne2020}, nutritional epidemiology \cite{Molitor2014}, occupational epidemiology \cite{Belloni2020} and genomic \cite{Rouanet2024}. In this paper, we focus on the use of BPRM models to deal with a censored survival health outcome as proposed by Belloni \textit{et al.} (2020) \cite{Belloni2020} and Liverani \textit{et al.} (2021) \cite{Liverani2021}. Compared to Belloni \textit{et al.} (2020) \cite{Belloni2020}, we assume an increasing continuous function to describe the instantaneous baseline hazard rate function, following Fendler \textit{et al.} (2024) \cite{Fendler2024}.
\newline

BPRM models are semi-supervised clustering models in which the prior number of clusters is assumed to be infinite: a Dirichlet process prior is assigned to the mixture weights. As a consequence, BPRM models belong to the class of Dirichlet Process Mixture (DPM) models. Standard MCMC algorithms are generally not suitable to infer mixture models like BPRM models. This is mainly due to both multiple modes in the joint posterior distribution, which requires the ability to efficiently visit these different modes, and to the unfixed number of non-empty clusters, which requires the ability to efficiently change the dimension of the approximate posterior distribution during the MCMC run. Thus, to infer BPRM models, Liverani \textit{et al.} (2015) \cite{Liverani2015} proposed an MCMC algorithm based on a slice sampler and including three label switching moves. Unfortunately, this advanced sampling scheme is not yet sufficiently effective : the estimated number of clusters may remain unstable, as reported in different applications \cite{Belloni2020, Rouanet2024, Giampiccolo2024}.  In this work, we extend Liverani \textit{et al.} (2015)'s algorithm with a parallel tempering algorithm, in order to  escape from local modes and thus, visit more efficiently the support of (potentially highly) multimodal posterior distributions. The parallel tempering algorithm for MCMC, also known as Metropolis-coupled MCMC or MC\textsuperscript{3}, allows a Markov chain to jump to a new highly probable position if stuck in a local mode \cite{Geyer1992}. Napier \textit{et al.} (2019) \cite{Napier2019} showed through a simulation study that this algorithm outperformed classic MCMC algorithms to recover a clustering partition from complex mixture models. The parallel tempering algorithm for MCMC is rarely used in the clustering literature and, to the best of our knowledge, it has never been used to infer BPRM models. In this paper, a simulation study is performed to compare the performance of the MCMC algorithm proposed in Liverani \textit{et al.} (2015) with an MCMC algorithm coupled with a parallel tempering algorithm, to infer BPRM models. We also propose to speed up the Bayesian inference of BPRM models using the JAX library in Python. \\
\newline
When fitting a BPRM model with an MCMC algorithm, the outputs are samples of the approximate posterior distribution of all the possible partitions\footnote{A partition corresponds to the assignment of the individuals into a specific number of clusters. For $n$ individuals, there exists $B_n = \sum_{k = 1}^{n} \frac{1}{k!}\sum_{j=0}^{k}(-1)^{k-j}\binom{k}{j}j^n$ different partitions.} of the individuals. Thus, different posterior numbers of non-empty clusters may be derived from these MCMC outputs. In order to  unravel a single clustering structure of the data, an optimal partition has to be selected via a post-processing step. Various methods have been proposed in the literature. Methods based on the minimisation of the distance to the mean similarity matrix and the Partitioning Around Medoids (PAM) algorithm are classically used in the BPRM literature \cite{Belloni2020, Molitor2010, Liverani2015, Giampiccolo2024}. Liverani \textit{et al.} (2015) \cite{Liverani2015} reports that PAM algorithm is less subject to Monte Carlo error. However, this post-treatment procedure is often criticized in the literature for being ad-hoc and not properly designed for DPM \cite{Wade2018}. Wade and Ghahramani (2018) \cite{Wade2018} proposed a new post-treatment procedure for DPM models based on decision theory. To the best of our knowledge, this post-treatment procedure has not yet been applied to BPRM models. In this work, we compare the performances of the three above post-treatment procedures via a simulation study. \\
\newline 
In the following, we first describe a BPRM model to deal with a censored survival health outcome as well as a mixture of categorical and continuous exposure covariates. Then, we describe our parallelised MCMC algorithm coupled with a parallel tempering algorithm to fit the BPRM model. In 
section 3, we recall the three methods of post-processing we compare to select an optimal partition of individuals. In section 4, we demonstrate the benefit of coupling an MCMC algorithm with parallel tempering to fit a BPRM model through a simulation study and compare the three methods of post-processing in terms of estimated number of non-empty clusters, proportion of misclassified individuals and relative bias on the risk parameters. In section 5, our methods are applied in radiation epidemiology to identify subgroups of uranium miners sharing similar risk of death by lung cancer (to estimate) and similar professional exposure profiles (to characterize) in the post-55 French cohort of uranium miners \cite{Rage2015}. Finally, a discussion is provided in the last section. 


\section{A BPRM model for censored survival health outcomes}

To facilitate the description of a BPRM model, it may be useful to divide it in three sub-models that must be specified and linked, through conditional independence assumptions:

\begin{itemize}
    \item The assignment sub-model : it defines non-empty clusters (or profiles) and assign each individual to one of these clusters ;
    \item The exposure sub-model:  it defines the probability distribution of the different exposure variables in each cluster, in order to characterize specific exposure profiles ;
    \item The outcome sub-model also called disease sub-model hereafter, when dealing with a censored survival health response : it defines the probability distribution of the outcome data (e.g., age at the occurrence of a given disease) in each cluster.
\end{itemize}

\subsection{\textit{The assignment sub-model}}

Let $C_i$ be the latent cluster label of individual $i$, $i \in \{1,...,n\}$ and $\phi = \{\phi_c, c \in \mathbb{N}^{*}\}$ the vector of assignment probabilities (or mixture weights) to each cluster $c$, such that $\phi_c=\mathbb{P}(C_i=c)$. The assignment sub-model defines the prior distribution of $\phi$ as a Dirichlet process based on the assumption that the prior number of clusters is infinite. Note that, as the datasets are finite in practice, only a finite posterior number of non-empty clusters can be estimated.

To assign a Dirichlet process prior on the vector $\phi$, we adopt the stick-breaking construction \cite{Pitman2002} defined as follows:
    \begin{equation}
     \begin{array}{ll}
      V_c | \alpha \sim Beta(1, \alpha) & \mbox{ i.i.d. for } c \in \mathbb{N}^*\\
         \phi_1 = V_1 &  \\
        \phi_c = V_c \prod_{l<c}(1 - V_l) & \mbox{ for } c \in \mathbb{N}^*\backslash\{1\}
    \end{array}
\end{equation}

$\alpha>0$ is an unknown parameter to estimate, called the concentration parameter. It indirectly defines the posterior number of non-empty clusters : the smaller $\alpha$ is, the smaller the posterior number of non-empty clusters is.

\subsection{\textit{The exposure sub-model}}


Positive continuous exposure variables are assumed to follow log-normal distributions and categorical exposure variables are assumed to follow categorical distributions. The unknown parameters defining these distributions are different for each cluster. Thus, they give a specific characterization to each cluster in terms of exposure profiles.


For an individual $i$ belonging to the cluster $C_i$, we make the following probabilistic modelling assumptions :
\begin{itemize}[noitemsep]
    \item the $k^{th}$ continuous exposure variable $X^k_i$ follows a log-normal distribution $\mathcal{LN}(\mu^k_{C_i}, \sigma^k_{C_i})$ for $k = \{1,...,K\}$ and $K$ the number of continuous exposure variables;
    \item the $j^{th}$ categorical exposure variable $X^j_i$ follows a categorical probability distribution $\mathcal{C}ategorical (p^j_{C_i})$ for $j = \{1,...,J\}$ and $J$ the number of categorical exposure variables. p$_c^j$ is the vector of  probabilities associated to the $M_{j}$ modalities of the exposure variable j within cluster $c$ with $\sum_{m=1}^{M_j}p_{c}^{j,m}=1$.
\end{itemize}
\subsection{\textit{The disease sub-model}}

In this work, we focus on the specific case of a censored survival health outcome.
Let $T_i$ be the age at the occurrence of the health event of interest of individual $i$, $i \in \{1,...n\}$, and $W_i$ the age at censorship. The observable random variables for the individual $i$ are : $Y_i = min(T_i, W_i)$ and $\delta_i$ the binary indicator of non-censorship ($\delta_i = 1$ if $T_i \leq W_i$, $\delta_i = 0$ if $T_i > W_i$).\\
\newline
The disease sub-model is a parametric survival model. The instantaneous hazard rate of presenting the health event at time $t$ for the individual $i$, noted $h_i(t)$, is defined by :
\begin{equation*}
    h_i(t) = h_0(t)g(\beta, C_i)
\end{equation*}
$h_0(t)$ is the instantaneous baseline hazard rate function. Following Fendler \textit{et al.} (2024) \cite{Fendler2024}, we assume the following increasing continuous function :
\begin{equation*}
h_0(t)=\xi t^{\nu -1}  \qquad \textrm{ with  } \xi>0,\nu>1
\end{equation*}
$g(\beta,C_i)$ is the hazard ratio function. As classically used in radiation epidemiology \cite{Belloni2020, Fendler2024, Hoffmann2017}, we propose to consider the following Excess Hazard Ratio (EHR) structure :
\begin{align}
g(\beta, C_i)=1 + \beta_{C_i}
\end{align}
with $\beta=\{\beta_c,c\in \mathbb{N}^{*}\}$ and $\beta_{c}\geq -1$  for all $c\in \mathbb{N}^{*}$, to ensure the positivity of $h_i(t)$. $\beta_c$ corresponds to the unknown health risk coefficient associated to cluster $c$.\\

\subsection{\textit{Writing the model in terms of a mixture model}}

\begin{figure}
    \centering
    \includegraphics[scale = 0.5]{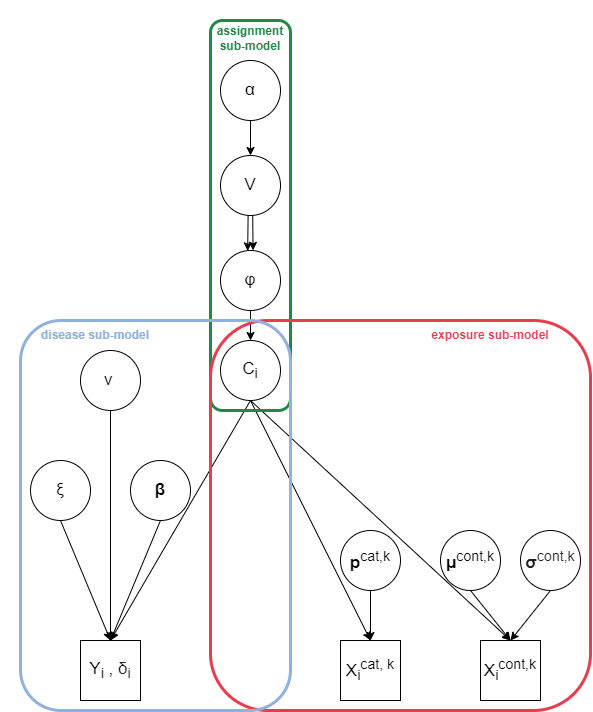}
    \caption{Directed Acyclic Graph (DAG) of the BPRM model. Circles indicate unknown quantities and rectangles indicate observed variables. Single
arrows indicate oriented probabilistic links between two quantities and double arrows indicate oriented deterministic links between two quantities.}
    \label{fig:DagBPRM}
\end{figure}

Combining the three sub-models (assignment, exposure and disease ones) gives the overall BPRM model whose Directed Acyclic Graph (DAG) is provided in Figure \ref{fig:DagBPRM}.
The contribution of the individual $i$, $i \in \{1,...,n\}$ to the likelihood can be written as follows: 
\begin{align*}
        \mathcal{L}(Y_i, \delta_i, X_i^{cont}, X_i^{cat} \mid \phi, \theta, \nu, \xi) & = \sum_{c = 1}^{+\infty} \phi_c [Y_i, \delta_i \mid  \beta_c, \nu, \xi] [X_i^{cont} \mid \mu_c, \sigma_c] \\
        & \qquad \times [X_i^{cat}\mid p_c]\\
    \end{align*}
where the notation $[.|.]$ defines a conditional density for continuous variables or a probability distribution for discrete variables. $\theta = \{\theta_c, c \in \mathbb{N}^{*}\}$ with $\theta_c=(\beta_c, \mu_c,\sigma_c,p_c)$, $\mu_c=(\mu_{c}^{1},\ldots,\mu_{c}^{K})$, $\sigma_c=(\sigma_{c}^{1},\ldots,\sigma_{c}^{K})$ and $p_c=(p_{c}^{1},\ldots,p_{c}^{
J})$.
The above equation allows to highlight the mixture structure of a BPRM model. 

\section{Bayesian inference}

\subsection{Prior distributions}
Working under the Bayesian paradigm requires specifying prior distributions on all unknown parameters. For the assignment sub-model, the strictly positive concentration parameter $\alpha$ follows a prior distribution $\mathcal{G}amma(2,1)$ as suggested by Liverani \textit{et al.} (2015)\cite{Liverani2015}. This prior ensures the posterior consistency of the posterior number of non-empty clusters\cite{Ascolani2023}. 
An other possibility could be to fix $\alpha$ but this choice leads to the posterior number of non-empty clusters being inconsistent \cite{GuhaHoNguyen2021}.\\
\newline
For the parameters of the exposure sub-model, we assume the following prior distributions :
\begin{itemize}
    \item $\mu^k_c \sim \mathcal{N}(0, 10^2)$, $k \in [\![ 1,K ]\!]$ and $c \in \mathbb{N}^*$
    \item $\sigma^k_c \sim \mathcal{G}(0.001,0.001)$,  $k \in [\![ 1,K ]\!]$ and $c \in \mathbb{N}^*$
    \item $p^j_c \sim \mathcal{D}irichlet (\frac{1}{2})$, $j \in [\![ 1,J ]\!]$ and $c \in \mathbb{N}^*$
\end{itemize}

For the disease sub-model, we consider the two following reparameterisations for the parameters defining the instantaneous baseline hazard function $h_0$:
\begin{itemize}
\item $\xi= \epsilon \times \tilde{\xi}$
\item $\nu'= \nu -1$
\end{itemize}
where $\epsilon$ is a scaling constant to be fixed. Following Fendler et al. (2024) \cite{Fendler2024}, we assign the following Gamma prior distributions on $\tilde{\xi}$ and $\nu'$: 
\begin{itemize}
    \item $\tilde{\xi} \sim \mathcal{G}(1,1)$;
    \item $\nu' \sim \mathcal{G}(0.001,0.001)$.
\end{itemize}

Finally, we assign a proper and informative prior distribution to each risk parameter $\beta_c$ for $c \in \mathbb{N}^{*}$ in order to help the inference. Indeed, as a risk parameter can be associated with an empty cluster, each $\beta_c$ must follow a proper prior distribution \cite{Molitor2010}. Thus, for any $c$, $\beta_c$ follows a $\mathcal{B}eta$ PERT distribution \cite{Vose1996}, which is a generalisation of the $\mathcal{B}eta$ distribution on a given interval. This distribution is characterised by the minimum and maximum bounds of its range of possible values and by its mode. In order to respect the constraint $\beta_c \geq -1$ for all c imposed by the EHR structure, the minimum bound of the $\mathcal{B}eta$ PERT distribution is always set to -1. The mode of the distribution is set to 0, corresponding to the case where the health risk is null. In order to correctly estimate the 95\% credible interval associated to each parameter $\beta_c$, we recommend setting the upper bound of its $\mathcal{B}eta$ PERT prior to at least three times the higher expected risk. 

\subsection{An advanced MCMC algorithm coupled with parallel tempering}

The Bayesian inference of our BPRM model is performed using an MCMC algorithm to sample from the joint posterior distribution of all
unknown parameters and latent class labels. We implemented in Python 3.8 the slice dependent Metropolis-Hastings sampler proposed by Walker (2007) \cite{Walker2007} for DPM models and adapted for BPRM models by Liverani \textit{et al.} (2015) \cite{Liverani2015} in a R package called "PReMiuM". Each unknown parameter of the vectors $\mu_c$, $\sigma_c$, $p_c$ and $V_c$ is updated with a Gibbs sampler as its associated full condition distribution is known. Finally, the concentration parameter $\alpha$ and the vector $\beta_c$ are updated with a Metropolis-Hastings sampler. For Metropolis-Hastings steps, we consider an adaptive phase, based on 100 steps of 100 iterations, to improve the convergence and the efficiency of the algorithm: the variance of each proposal distribution is updated to target an acceptance rate of 40\% for single parameters and 20\% for vectors. 

When using an MCMC algorithm to fit a mixture model, a well-known problem is label-switching : the likelihood is symmetric (i.e., the model is not identifiable) and invariant under relabelling of the mixture components \cite{Stephens2000}. Moreover, the stick-breaking construction of the Dirichlet process considered in the BPRM does not carry this property, leading to the posterior distribution of the algorithm being influenced to the arbitrary starting values of the cluster labels. In order to ensure that the MCMC algorithm explores the space of all probable permutations, the three label-switching moves recommended by Hastie, Liverani and Richardson (2015) \cite{HastieLiveraniRichardson2015} were implemented.
\newline

\begin{figure}
    \centering
\includegraphics[width=0.7\linewidth]{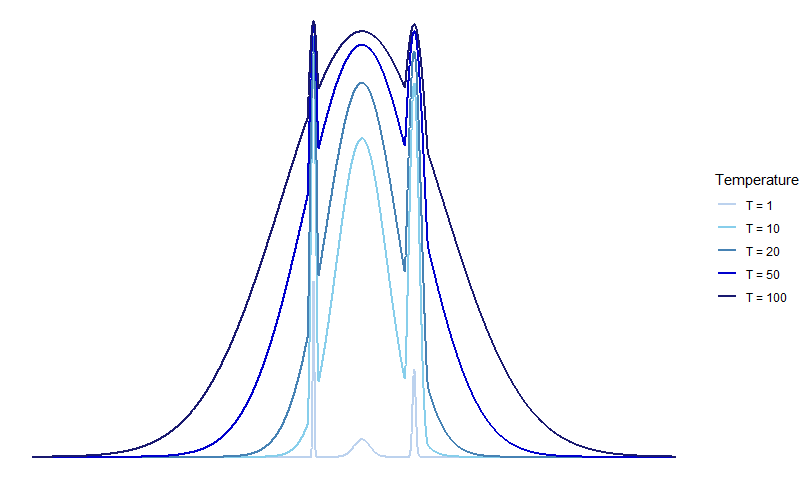}
    \caption{Impact of raising a multimodal function to a power of $\frac{1}{T}$ for different temperature values $T \in \{1,10, 20,50,100\}$}
    \label{fig:parallel_temp}
\end{figure}

The joint posterior distribution of a BPRM  model is often multimodal \cite{Belloni2020}. 
In order to help the Markov chains escape from local modes, we have coupled the previously described MCMC algorithm with a parallel tempering (PT) algorithm. 
The general principle of PT is to smooth the likelihood function, and thus, the targeted posterior distribution, by raising it to a power of $\frac{1}{T}$ where $T \in [1; + \infty]$ is a fixed parameter called temperature \cite{Sambridge2014}. Figure \ref{fig:parallel_temp} illustrates that a multimodal function tends to be smoothed by high temperature values.
\newline

Let $D_i=(Y_i, \delta_i, X_i^{cont}, X_i^{cat})$ be the set of all observable random variables, $\boldsymbol{C}=\{C_i, i\in\{1,\ldots,n\}\}$ the vector of latent class labels and $\omega=(\boldsymbol{C}, \phi, \theta, \nu, \xi)$ the set of all unknown parameters and latent class labels of our BPRM model. For a fixed temperature $T$, the target distribution in which values of $\omega$ must be sampled is proportional to $\prod_{i=1}^n[D_i|\omega]^{\frac{1}{T}}[\omega]$. $L$ Markov chains are run in parallel at $L$ different values of temperature,  including the temperature $T = 1$. For $l \in [\![1,L]\!]$, we note $\omega_l^t$ the position of the Markov chain at temperature $T_l$ at iteration $t$ of the MCMC algorithm. Every $n_{pt}$ iterations, $\omega_{l}^t$ and $\omega_{l +1}^t$ are exchanged with probability $p_{pt}$, where $l$ is drawn randomly from $\{1, ..., L -1 \}$, and 
\begin{align*}
    p_{pt} &= \min \Big(1, \prod_{i = 1}^n\frac{[D_i|\omega_{l+1}]^{\frac{1}{T_{l}}}[\omega_{l +1}][D_i|\omega_{l}]^{\frac{1}{T_{l +1}}}[\omega_{l}]}{[D_i|\omega_{l+1}]^{\frac{1}{T_{l +1}}}[\omega_{l +1}][D_i|\omega_{l}]^{\frac{1}{T_{l}}}[\omega_{l}]}\Big)\\
&= \min\Big(1, \big(\prod_{i = 1}^n\frac{[D_i|\omega_{l+1}]}{[D_i|\omega_{l}]}\big)^{T_{l + 1}- T_l} \Big)
\end{align*}
The choice of the set of temperatures and of $n_{pt}$ can be automatically adjusted at the start of the algorithm \cite{EarlDeem2005}, but this is out of the scope of this study. Nevertheless, the set of temperature values must be chosen so that the acceptance rate of the exchange is approximately 20\% and choose $n_{pt}$ sufficiently large so that the chains have time to converge to their respective target distribution after each exchange \cite{Laloy2016}. 
Taking into account these recommendations, we propose the following set of temperatures $T \in \{1,2,5,10, 20\}$ and $n_{pt} = 1000$.\\
\newline

\subsection{Accelerating the inference algorithm}
The time complexity of our MCMC algorithm coupled with a parallel tempering algorithm mainly depends on the amount of computer time it takes to update the vector of latent class labels $\boldsymbol{C}$ at each iteration. In order to speed up inference, we propose to simultaneously update all the components of the vector $\boldsymbol{C}$. This task of the algorithm is embarrassingly parallel. Note that no update of the cluster-specific parameters i.e., $\phi_c$ and $\theta_c=(\beta_c,\mu_c,\sigma_c,p_c)$ for $c \in \mathbb{N}^{*}$ is done during the update of $\boldsymbol{C}$. We have implemented the parallelisation of this task in Python using the recent and powerful library JAX\cite{jax2018github} for high-performance numerical computing. Interestingly, this has reduced the calculation time on our CPU-only computer from an average of $4.9$ seconds per iteration of the MCMC algorithm to $0.30$ seconds. This means a $94$\% reduction in calculation time per iteration.
 \\

The Python code of our parallelised MCMC algorithm coupled with a parallel tempering algorithm is available at \hyperlink{https://github.com/fendlerj/BPRM_PT}{https://github.com/fendlerj/BPRM\_PT}.

\section{Post-processing}

In the following, we denote $P^{(t)}$ the partition of $n$ individuals at iteration $t$ of the MCMC algorithm and $P^*$ the optimal partition to choose. Furthermore, we denote $S^{(t)}$ the similarity matrix at iteration $t$, defined by: $$(S^{(t)})_{i,j} = \left\{
    \begin{array}{ll}
        1 & \mbox{if individuals }i\mbox{ and }j\mbox{ are assigned to the same cluster}  \\
        0 & \mbox{if not.}
    \end{array}
\right.$$ and $\hat{S}$ the average of the similarity matrices over all the iterations, i.e. $\hat{S} = \frac{1}{n_{iter}} \sum_{t=1}^{n_{iter}}S^{(t)}$ 
where $n_{iter}$ is the number of iterations of the posterior sample derived from the MCMC algorithm (i.e., excluding the burnin step). Thus, $\hat{S}_{i,j}$ is the proportion of iterations in which individuals $i$ and $j$
are assigned into the same cluster among the $n_{iter}$ iterations.\\

\subsection{Minimisation of the distance to the mean similarity matrix (BL)} 

This first method we consider in this work consists in choosing the optimal partition $P*$ among those sampled during the MCMC run and whose associated similarity matrix minimises the distance, in the Frobenius norm, to the mean similarity matrix $\hat{S}$. More formally, 
$P^*$ is the partition $P^{(t^*)}$ obtained at iteration $t^*$ with : $$t^* = {\arg\,\min}_{t \in \{1,...,n_{iter}\}} \sum_i \sum_j ((\hat{S})_{i,j} - (S_t)_{i,j})^2.$$ 
This method is equivalent to choosing the partition that minimises the posterior expectation of the Binder loss function (BL)\cite{Liverani2015, FritschIckstadt2009}. 

\subsection{Partitioning around medoids (PAM) algorithm}

The partitioning around medoids (PAM) algorithm \cite{KaufmanRousseeuw1987} is applied to the average dissimilarity matrix $\hat{M} = (1 - \hat{S})$ for a fixed number of clusters $k$ and a partition $P_k$ is derived for $k$ varying between $1$ and $K$. $P^*$ is defined as the partition $P_k*$ such that : $$k^* = \arg \min_{k \in {1,...,K}} s_k$$ where $s_k$ is the silhouette of the partition in $k$ clusters, $k \in \{1,..., K\}$. This algorithm is fully deterministic. 

\subsection{Minimising the variation of information (VI)}
This method, based on decision theory, was introduced by Wade and Ghahramani (2018)\cite{Wade2018}. The variation of information $VI$ between two partitions $P$ and $\hat{P}$ is defined by \cite{Meila2007} :
\begin{align*}
VI(P, \hat{P}) &= - \sum_{i=1}^K \frac{n_{i+}}{n}\log_2(\frac{n_{i+}}{n}) - \sum_{j=1}^{\hat{K}} \frac{n_{+j}}{n} \log_2(\frac{n_{+j}}{n}) \\
&\:\:\:+ 2 \sum_{i=1}^K \sum_{j=1}^{\hat{K}} \frac{n_{ij}}{n} \log_2(\frac{n_{ij}n}{n_{i+}n_{+j}})
\end{align*}
where $K$ (resp. $\hat{K}$) is the number of clusters in the partition $P$ (resp. $\hat{P}$), $n_{ij}$ is the sample size corresponding to the intersection of the cluster $i$ of $P$ and the cluster $j$ of $\hat{P}$, $n_{i+} = \sum_j n_{ij}$ and $n_{+j} = \sum_i n_{ij}$. \\

This function quantifies the difference in information (measured using the entropy function) between two partitions $P$ and $\hat{P}$ of $n$ individuals. For each of the two partitions, the 
difference is calculated between the information contained within each cluster and the information contained in the intersection of the two partitions. $P^*$ is defined as the partition minimising the posterior expectation of the cost function $VI$, i.e.: $$P^{*} = \underset{\hat{P}}{\operatorname{argmin }} \mathbb{ E}[VI(P, \hat{P})|\boldsymbol{D}_{1:n}]$$ with $\boldsymbol{D}_{1:n}=\{D_i; i\in\{1,\ldots,n\}\}$. 
In practice, the above minimisation is applied to the partitions resulting from the MCMC algorithm with complete linkage between the clusters (the distance between two clusters is the distance between the two furthest points of the clusters) using the R package \textit{mcclust.ext} \cite{mcclust.ext}. 

\section{Simulation study}

\subsection{Aim and performance criteria}

A simulation study is designed to compare the performances of :
\begin{itemize}
\item two learning algorithms to infer BPRM models : the MCMC algorithm proposed by Liverani et al. (2015) (MCMC hereafter) and our MCMC algorithm coupled with a parallel tempering algorithm  (MCMC \& PT hereafter);
\item three post-processing procedures : BL, PAM and VI.
\end{itemize}

Four performance criteria are considered:
\begin{itemize}
\item the estimated number of non-empty clusters to assess the ability of each method to recover the true simulated number of non-empty clusters;
\item the proportions of individuals either classified as at risk when they are not ($\beta = 0$), or classified as without risk when they are ($\beta >0$); 
\item the posterior mean of each cluster-specific parameter;
\item the (relative or absolute) bias in estimating the $\beta$ excess risk coefficient but only to compare the learning algorithms. 
\end{itemize} 

\subsection{Simulation scenarios}

Four simulation scenarios called $S_1$, $S_2$, $S_3$ and $S_4$ are considered. For each scenario, four clusters of individuals, called A, B, C and D, are built, one of which is composed of individuals that are non exposed. Three continuous exposures covariates ($X_1$, $X_2$ ans $X_3$) and one additional continuous covariate ($X_4$) are considered. 2000 individuals are equally distributed between these four clusters (i.e., 500 individuals per cluster). A censored survival health outcome is generated for each individual, according to the previously described disease sub-model. Guided by the case study in radiation epidemiology described in the next section, we consider the following parameter values for the instantaneous baseline risk function : $\xi=5\times 10^{-25}$ and $\nu= 5$. Four continuous exposure variables are assigned to each individual and generated according to the exposure sub-model previously described. They follow a lognormal distribution with a cluster-specific mean value ($\mu^k$ for the k-th variable) and a standard deviation value ($\sigma^k$ for the k-th variable), both on logarithmic scale. Finally, 100 data sets are generated for each scenario.\newline

The cluster-specific parameter values fixed for each simulation scenario are given in Table \ref{tab:simu4}. Regarding scenarios S1, S2 and S3, each of the exposed clusters B and C is associated to a different excess risk value $\beta$. The greatest value is systematically assigned to cluster $C$. Moreover, the mean values $\mu^k$ (at logarithmic scale) of the four exposure variables are equal between these three scenarios but not their standard deviation values $\sigma^k$ (at logarithmic scale). The further apart the risk coefficient values $\beta$ or the smaller the standard deviation values $\sigma^k$, the better the separation between the clusters. Thus, scenario S2 leads to the most separated clusters whereas scenario S3 corresponds to the closest clusters and thus, to the most difficult to identify. Scenario S1 can be seen as an intermediate case.\newline

 Regarding the last scenario S4, the two exposed clusters B and C are associated to the same excess risk value $\beta=3$. Moreover, they differ only by the mean of the fourth exposure variable and the value of the 
 This scenario allows to consider the case where two clusters with high excess risk exist but are not associated to the same exposure profiles.  \\

\begin{table}
    \centering
    \resizebox{\textwidth}{!}{
    \begin{tabular}{|c||c||c|c|c|c|c|c|c|c|c|}
    \hline
    Scenario & Cluster & $\beta$ & 
    $\mu^1$ & $\mu^2$ & $\mu^3$ & $\mu^4$ & $\sigma^1$ & $\sigma^2$ & $\sigma^3$ & $\sigma^4$\\
    \hline\hline
    \multirow{4}{*}{S1} & A & 0 & 1.41 & 0.74 & 6.9 & 3.31 & 0.81 & 0.57 & 0.46 & 0.22 \\
    & B & 2.5 & 3.09 & 1.89 & 7.54 & 3.27 & 0.37 & 0.40 & 0.36 & 0.19 \\
    & C & 5 & 4.18 & 2.92 & 8.17 & 3.33 & 0.33 & 0.38 & 0.33 & 0.19 \\
    & D & 0 & - & - & - & 3.30 & - & - & - & 0.19\\
    \hline
    \multirow{4}{*}{S2} & A & 0 & 1.41 & 0.74 & 6.9 & 3.31 & 0.24 & 0.17 & 0.14 & 0.07 \\
    & B & 2.5 & 3.09 & 1.89 & 7.54 & 3.27 & 0.11 & 0.12 & 0.11 & 0.06 \\
    & C & 5 & 4.18 & 2.92 & 8.17 & 3.33 & 0.10 & 0.11 & 0.10 & 0.06 \\
    & D & 0 & - & - & - & 3.30 & - & - & - & 0.06\\
    \hline
    \multirow{4}{*}{S3} & A & 0 & 1.41 & 0.74 & 6.9 & 3.31 & 1.38 & 0.97 & 0.78 & 0.37 \\
    & B & 1.5 & 3.09 & 1.89 & 7.54 & 3.27 & 0.63 & 0.68 & 0.61 & 0.32 \\
    & C & 3 & 4.18 & 2.92 & 8.17 & 3.33 & 0.56 & 0.65 & 0.56 & 0.32 \\
    & D & 0 & - & - & - & 3.30 & - & - & - & 0.32 \\
    \hline
    \multirow{4}{*}{S4} & A & 0 & 1.41 & 0.74 & 6.9 & 3.31 & 1.38 & 0.97 & 0.78 & 0.37 \\
    & B & 3 & 4.18 & 2.92 & 8.17 & 3.22 & 0.63 & 0.68 & 0.61 & 0.32 \\
    & C & 3 & 4.18 & 2.92 & 8.17 & 3.33 & 0.56 & 0.65 & 0.56 & 0.32 \\
    & D & 0 & - & - & - & 3.30 & - & - & - & 0.32 \\
    \hline
    \end{tabular}
    }
    \caption{Cluster-specific parameter values fixed for each of the four simulation scenarios : $\beta$ is the excess risk coefficient, $\mu^k$ and $\sigma^k$ are the mean and standard deviation - at logarithmic scale - of the k-th exposure variable respectively. }
    \label{tab:simu4}
\end{table}

\subsection{Results}

Table \ref{tab:nb_clust} gives some summary statistics of the empirical distribution of the posterior number of non-empty clusters estimated from 100 datasets generated for each simulation scenario with each post-processing method and each learning algorithm. Scenario S2 is the scenario for which the clusters are the most separated. The variances of the exposure variables are the lowest and the excess risk coefficients are very different. In this case, the three post-processing methods, applied to the outputs of the MCMC or MCMC \& PT algorithm, correctly identify the four generated clusters. In addition, the three post-processing procedures appear to be stable. Scenario S1 generates clusters that are slightly less different. Indeed, although the excess risk coefficients are identical to those in scenario S2, the variances of the exposure variables are higher. In this context, after running the classical MCMC algorithm, the PAM algorithm identifies the four generated clusters slightly better than when minimising the distance to the mean similarity matrix (BL) or minimising the variation of information (VI). We can see that the latter two post-processing methods tend to underestimate the number of non-empty clusters. However, the performance of the post-processing BL, VI and PAM is equivalent for this scenario S1 when applied to the outputs of our proposed
MCMC \& PT algorithm. As previously described, scenarios S3 and S4 generate clusters that are less well separated than in S1. 
In this context, the number of non-empty clusters is underestimated by the post-processing BL and VI whether using the outputs of the classical MCMC or MCMC \& PT algorithm. Interestingly, the PAM algorithm tends to find the true number of non-empty clusters whether using the outputs of the classical MCMC or MCMC \& PT algorithm. However, we can see that it is then less stable than in S1 and S2 : the first and last quartile of the number of non-empty clusters estimated over 100 datasets are equal to 3 and 5 respectively.\newline

\begin{figure}
    \centering
    \includegraphics[width=1\linewidth]{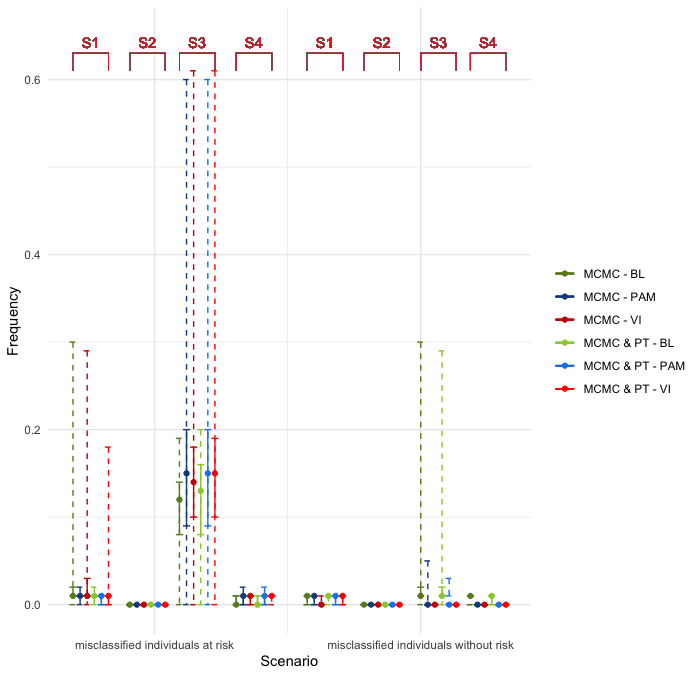}
    \caption{Comparison of the proportions of misclassified individuals  estimated over 100 simulated datasets when applying each post-processing procedure on the outputs of the MCMC or MCMC \& PT algorithm, for the 4 simulation scenarios. The points denote the means of the empirical distributions. The solid lines define the 75\% uncertainty intervals and the dotted lines define the 95\% uncertainty intervals.}
    \label{fig:mal_classé_4}
\end{figure}

\begin{table}
    \centering
    \begin{tabular}{|c||c|c|c||c|c|c|}
      \hline
      Scenario & \multicolumn{3}{c|}{MCMC} &  \multicolumn{3}{c|}{MCMC \& PT} \\
      \hline
      & BL & VI & PAM & BL & VI & PAM \\
      \hline
      S1 & 3.83 & 3.73 & 4.00 & 4.01 & 3.95 & 4.00 \\
      & (4-4) &  (3-4) & (4-4) & (4-4) & (4-4) & (4-4) \\
      \hline
      S2 & 4.00 & 4.00 & 4.03 & 4.00 & 4.00 & 4.01\\
      & (4-4) & (4-4) & (4-4) & (4-4) & (4-4) & (4-4)  \\
      \hline
      S3& 3.22 & 3.00 & 3.86 & 3.25 & 3.00 & 3.79 \\
      & (3-3) & (3-3) & (3-5) & (3-3) & (3-3) & (3-3) \\
      \hline
      S4 & 3.09 & 3.00 & 4.05 & 3.07 & 3.00 & 3.95\\
      & (3-3) & (3-3) & (3-5) & (3-3) & (3-3) & (3-5) \\
      \hline
    \end{tabular}
    \caption{Mean (first-last quartile) of the empirical distribution of the posterior number of non-empty clusters estimated from 100 datasets generated for each simulation scenario, from the classical MCMC algorithm (MCMC) or MCMC algorithm coupled with parallel tempering (MCMC \& PT) when applying each post-processing procedure.}
    \label{tab:nb_clust}
\end{table}
Figure \ref{fig:mal_classé_4} presents the proportions of  misclassified individuals at risk (left-hand side of the figure) and of misclassified individuals without risk (right-hand side of the figure), estimated over 100 simulated datasets when applying each post-processing procedure on the outputs of the MCMC or MCMC \& PT algorithm, for the 4 simulations scenarios.
As expected, the proportions of misclassified individuals for scenario S2 are very low for all procedures. It is interesting to see that for S1 (less separated clusters than S2) the proportions of individuals that are without risk but classified as at risk remain very low. On the other hand, under scenario S1, the following procedures - MCMC with BL and VI and MCMC \& PT with VI - can lead to relatively high proportions of individuals that are at risk but classified without risk. Interestingly, an improvement can be noticed when using the outputs of our MCMC algorithm coupled with a parallel tempering algorithm. Under scenario S3, for which the generated clusters are the closest and thus, the most difficult to identify, the post-processing BL leads to higher proportions of misclassified individuals that are without risk but classified as at risk than PAM and VI. However, all the post-processing procedures classify the at-risk individuals fairly poorly. All in all, the PAM and VI post-processing procedures seem to offer a better compromise under scenario S3. Finally, as expected, the scenario S4 leads to very few misclassified individuals given that individuals are either without risk (i.e., $\beta$ =0) or at very high risk (i.e., $\beta=3$).\newline

\begin{figure}
    \centering
    \includegraphics[width=1\linewidth]{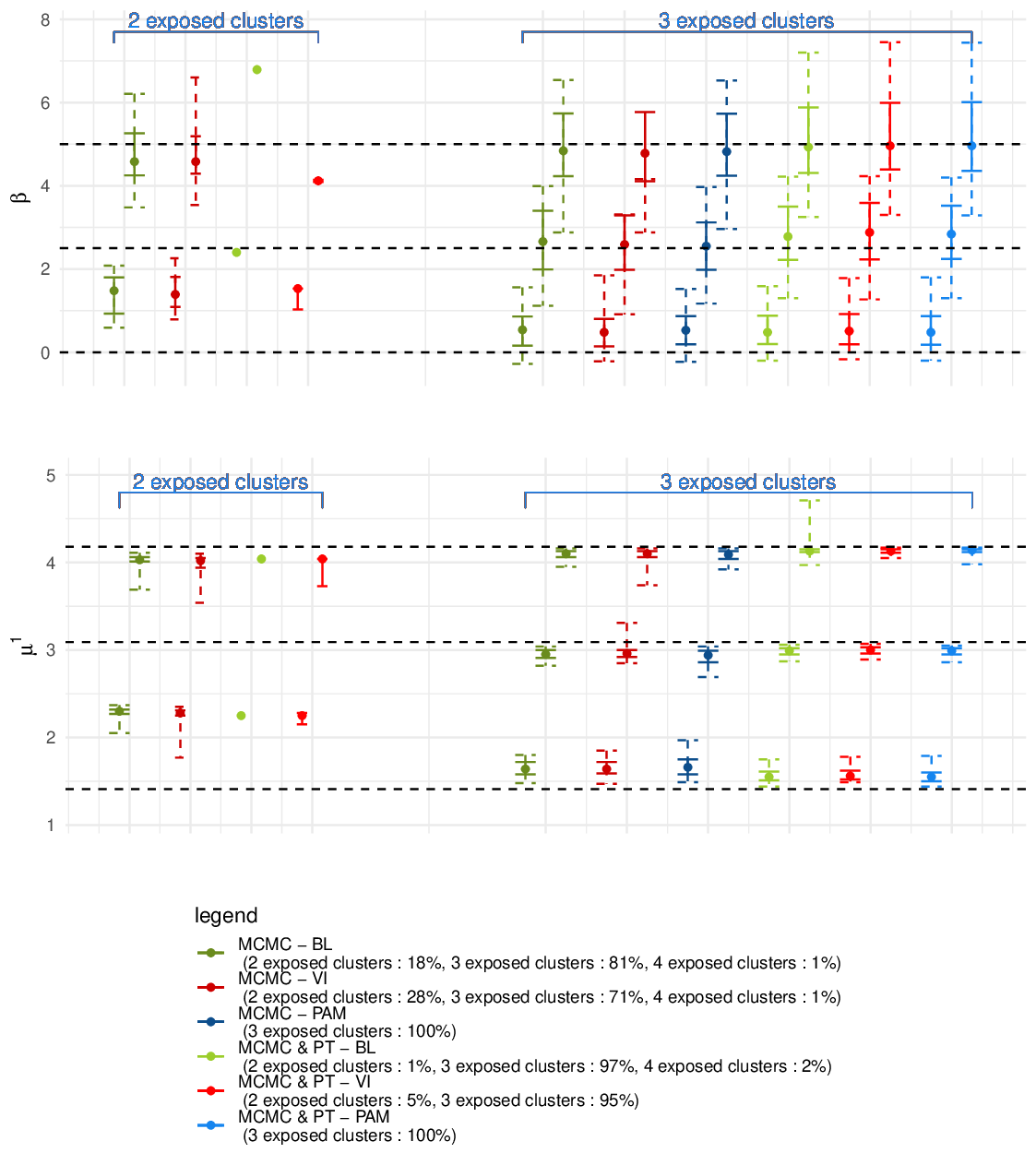}
    \caption{Means (points), 75\% (solid lines) and 95\% (dashed lines) credible intervals of the 100 posterior means estimated for $\beta$ (at the top) and $\mu^1$ (at the bottom) in each estimated exposed cluster, for datasets simulated under the scenario S1 and given the estimated number of exposed clusters. The horizontal dashed lines correspond to the true values of the parameters for the three true exposed clusters.}
    \label{fig:mean_param_beta_mu1}
\end{figure}

Table \ref{tab:biais_4gp} provides the mean as well as the first and last quartile of the empirical distribution of the relative biases associated to the estimation of the $\beta$ excess risk parameters (or absolute biases for $\beta_A$ that is equal to zero) and estimated over 100 datasets, for each learning algorithm (MCMC or MCMC \& PT) and under the four simulation scenarios. A relative bias (RB) is first calculated for each individual $i$ and each iteration $t$ of a posterior sample as : $$RB_{i,t} = \frac{\beta_{C_i}^{(t)} - \beta_{true, i}}{\beta_{true, i}}$$ where $\beta_{true, i}$ is the true value of $\beta$ for individual $i$ and $\beta_{C_i}^{(t)}$ is the estimated value of $\beta$ for the cluster $C_i$ to which individual $i$ is assigned at iteration $t$.
This relative bias is then averaged over all the iterations of the posterior sample and all the individuals belonging to the same cluster.
The fact that this bias is calculated at the individual level only makes it possible to compare the two learning algorithms (MCMC and MCMC \& PT). For the four simulation scenarios, the relative (or absolute) biases are similar between the MCMC and MCMC \& PT algorithms.  \\
\begin{table}
    \centering
    \resizebox{\textwidth}{!}{
    \begin{tabular}{|c||c|c|c||c|c|c|}
        \hline
        & \multicolumn{3}{c||}{MCMC}  & \multicolumn{3}{c|}{MCMC \& PT}\\
        \hline 
        & $\beta_A$ & $\beta_B$ &$\beta_C$ & $\beta_A$ & $\beta_B$ &$\beta_C$ \\
        \hline
       S1 & 0.56 & 0.02 & -0.02 & 0.60 & 0.12 & 0.03 \\
       & 0.21; 0.89 & -0.21;0.24 & -0.15; 0.15 & 0.18;0.95 & -0.09; 0.40 & -0.13;0.22\\
       \hline
       S2 & 0.20 & 0.12 & 0.04 & 0.21 & 0.14 & 0.05 \\
       & -0.07;0.41 & -0.13;0.32 & -0.07;0.17 & -0.08;0.44 & -0.12;0.35 & -0.05;0.18\\
       \hline
       S3 & 0.62 & 0.24 & -0.04 & 0.73 & 0.26 & -0.04\\
       & 0.31;0.88 & -0.05 ; 0.58 & -0.23 ;0.16 & 0.37 ;1.06 & -0.00;0.56 & -0.19;0.13\\
       \hline
        S4 & 0.16 & 0.02 & 0.02 & 0.23 & 0.03 & 0.03 \\
        & -0.04;0.36 & -0.13;0.21 & -0.14;0.21 & -0.02;0.41 & -0.11;0.22 &  -0.11;0.22\\
        \hline
    \end{tabular}
    }
    \caption{Mean (first and last quartile) relative (resp. absolute) biases on $\beta_B$ et $\beta_C$ (resp. $\beta_A$)}
    \label{tab:biais_4gp}
\end{table}
\newline
Figure \ref{fig:mean_param_beta_mu1} shows the means as well as the 75\% (solid lines) and 95\% (dashed lines) credible intervals of the 100 posterior means estimated for $\beta$ and $\mu^1$ in each cluster of exposed individuals for datasets simulated under the scenario S1 and given the number of clusters of exposed individuals.  Results for the parameters $\mu^2$, $\mu^3$ and $\mu^4$ are given in supplementary materials. The PAM post-processing always leads to estimate the true number of clusters of exposed individuals, equal to 3. The post-processing BL and VI applied to the outputs of the classical MCMC algorithm do not allow to estimate the true number of exposed clusters for 19\% and 29\% of the datasets respectively. Basically, when two exposed clusters are estimated, the individuals of cluster B with the lowest true strictly positive excess risk equal to 2.5 tend to be gathered with the individuals of cluster A for which the true $\beta$ equal to $0$. However, the above proportions of error are reduced to 3\% for BL and 5\% for VI when the Bayesian inference is performed with our MCMC \& PT algorithm. For all the post-processing methods, when the true number of clusters of exposed individuals is correctly estimated, the posterior mean of $\beta$, $\mu^1$, $\mu^2$, $\mu^3$ and $\mu^4$  are relatively well estimated. The estimation uncertainty is higher for the parameter $\beta$ than for $\mu^1$, $\mu^2$, $\mu^3$ and $\mu^4$. This is most probably due to the potentially high level of censorship on survival health outcomes that inform the excess risk coefficient $\beta$. The standard deviation parameters $\sigma^1$, $\sigma^2$, $\sigma^3$ and $\sigma^4$ are slightly overestimated (results not shown).\newline

Scenario S1 is chosen to illustrate the results because, on the one hand, the generated clusters overlap sufficiently so that the learning algorithms and post-processing methods can fail to recover the true partition of individuals for some simulated datasets, and on the other hand, the generated clusters are sufficiently separated so that the learning algorithms and post-processing methods can also recover the true partition of individuals for some datasets. 
However, note that we find similar results for scenario S2. Scenario S3, for which the generated clusters overlap too much to be easily identified, leads to estimate 2, 3 or 4 clusters of exposed individuals. In that case, the VI post-processing systematically identifies two exposed clusters (instead of 3). BL leads to estimate two or three exposed clusters. However, when three exposed clusters are estimated, two of them have  identical posterior distributions for $\beta$, $\mu_1$, $\mu_2$ and $\mu_3$. The PAM post-processing estimates two, three or four exposed clusters. When three exposed clusters are found, one of them is without risk, one of them is with a moderate risk and the last one has a high risk. The number of exposed clusters as well as the posterior mean of the cluster-specific parameters are better estimated when the MCMC algorithm is coupled with the PT algorithm. Finally, regarding the last scenario S4, none of the post-processing methods we have considered in this work is able to recover the 3 exposed clusters. VI and BL systematically identify two exposed clusters when the PAM algorithm estimates either 2 or 4 exposed clusters with a probability around 50\% for both cases. The results obtained for scenario S2, S3 and S4 are given in supplementary materials.

\section{Application to the post-55 French cohort of uranium miners}

The population of uranium miners is a reference population for studying the health effects of chronic exposure to radon and its short-lived decay products (referred to as radon hereafter). However, these nuclear workers are also chronically exposed to gamma rays and uranium dusts, but the health effects induced by a co-exposure to these different sources of ionizing radiation (IR) are not known : they are most often studied source by source \cite{Vacquier2011}.
Recently, Belloni \textit{et al.} (2020) \cite{Belloni2020} have estimated the risk of lung cancer death associated with chronic exposure to multiple sources of IR in the post-55 French cohort of uranium miners (end of follow-up: December, 31st 2007), using a Restricted Profile Regression Mixture (RPRM) model. Unlike BPRM, this kind of model requires one to fix the number of non-empty clusters. This choice has been made as a consequence of the poor convergence of the MCMC algorithm originally implemented to fit a BPRM model, especially with respect to the estimated number of non-empty clusters. In this work, we have applied our improved MCMC algorithm on the last update of the French cohort of uranium miners (end of follow-up: December, 31st 2014) allowing the estimate of the number of non-empty clusters as well.

\subsection{Data description}

The post-55 French cohort of uranium miners gathers 3377 males who were employed as uranium miners for at least 1 year in the CEA-COGEMA group between 1955 and 1990. More details about the methods of data collection (e.g., vital status, causes of death, . . .) have been described previously \cite{Rage2015}. In this work, we have only considered 3376 uranium miners: one miner has been excluded due to several abnormally high values of exposure to uranium dust. \newline

Table \ref{tab:StatutsVitaux_post55} provides the vital status at the end of follow-up : December 31\textsuperscript{st}, 2014. At this date, 2188 (64.8\%) of the miners are still alive. 130 (3.9 \%) are deceased of lung cancer. Only 20 (0.6\%) are lost to follow-up.\newline
\newline
\begin{table}[]
    \centering
    \begin{tabular}{lc}
    \hline
         Vital status & N (\%)  \\
         \hline
        Alive & 2188 (64.8)\\
        Deceased (Lung cancer) & 130 (3.9)\\
        Deceased (Other causes) &  1038 (30.7)\\
        Lost to follow-up & 20 (0.6) \\
        \hline
    \end{tabular}
    \caption{Vital status at december 31\textsuperscript{st}, 2014 in the post-55 French cohort of uranium miners (N = 3376 miners)}
    \label{tab:StatutsVitaux_post55}
\end{table}

In the cohort, the annual exposure to radon, gamma rays and uranium dusts is provided for all the workers. The main exposure characteristics are recorded in Table \ref{tab:stat_desc_post55}. In this work, we have considered the cumulative exposure to occupational radon in Working Level Month (WLM), gamma rays in miliSivierts (mSv) and uranium dust in becquerel hour per cubic mete (Bq.m\textsuperscript{-3}.h) during the whole follow-up period of each miner. A five year lagged is applied to the exposures to account for the delay of cancer occurence \cite{UNSCEAR2006}. An hypothesis of survival model is that the covariates are constant over time, we have left-truncated the follow-up of each uranium miner to his date of last exposure. Indeed, we consider that the time a miner enters in the cohort is the time of last exposure.  
\begin{table}[]
    \centering
    \resizebox{\textwidth}{!}{
    \begin{tabular}{lccc}
    \hline
         & Mean & Standard deviation & (Min-Max) \\
         \hline
        Age at entry in the study (in years) & 28.3 & 7.0 & (16.9 ; 57.7) \\
        Age at the end of follow-up (in years) & 65.7 & 11.4 & (19.5 ; 85.0) \\
        Duration of the follow-up (in years) & 37.4 & 11.1 & (0.1 ; 58.0) \\
        Duration of employment (in years) & 17.2 & 10.53 & (1.0 ; 57.6) \\
        Cumulated exposure to radon (in WLM) & 17.8 & 24.8 & (0.0 ; 128.4) \\
        Duration of the radon exposure (in years) & 12.9 & 8.5 & (1.0 ; 41.0) \\
        Age at first radon exposure (in years) & 29.2 & 7.5 & (16.7 ; 64.0) \\
        Cumulated exposure to gamma rays (in mSv) & 54.9 & 71.0 & (0.2 ; 470.0) \\
        Duration of the exposure to gamma rays (in years) & 13.3 & 8.7 & (1.0 ; 41.0) \\
        Age at first exposure to gamma rays (in years) & 29.1 & 7.5 & (16.5 ; 59.8) \\
        Cumulated exposure to uranium dusts (in Bq.m\textsuperscript{-3}.h) & 1641.0 & 1428.4 & (0.4 ; 10355.9) \\
        Duration of the exposure to uranium dusts (in years) & 13.0 & 8.0 & (1.0 ; 39.0)\\
        Age at first exposure to uranium dusts(en years) & 29.1 & 7.5 & (16.5 ; 59.8) \\
        \hline
    \end{tabular}
    }
    \caption{Main characteristics of the occupational exposure to ionizing radiation in the post-$55$ French cohort of uranium miners (N=3376 miners) at december, 31\textsuperscript{st} 2014}
    \label{tab:stat_desc_post55}
\end{table}

\subsection{Details about the exposure variables}
In this work, five continuous and two categorical exposure variables are considered for each uranium miner $i$:
\begin{itemize}
    \item Cumulative exposure to occupational radon $X^R_i$ ;
    \item Cumulative exposure to occupational gamma rays $X^G_i$ ;
    \item Cumulative exposure to occupational uranium dust $X^P_i$ ;
    \item Minimum age at first exposure $X^A_i$ to one of the three radiological sources of interest in years ; 
    \item Maximum duration of exposure $X^D_i$ when considering the three radiological sources of interest in years ; 
    \item Job type $X^J_i$ (with 5 modalities) most occupied by miner $i$ :
    \begin{itemize}
        \item Hewer before mechanization ;
        \item Hewer after mechanization ;
        \item Other underground work before mechanization ;
        \item Other underground work after mechanization ;
        \item Surface work;
    \end{itemize}
    \item Localization of the mine $X^M_i$ most occupied by miner $i$ with 2 modalities based on the type of deposit :
    \begin{itemize}
        \item Herault ;
        \item Other localization ;
    \end{itemize}
    
\end{itemize}

As not all the miners are exposed to the three studied sources of ionizing radiation, binary indicators to be at least one year occupationally exposed to radon $X^{IR}_i$, gamma rays $X^{IG}_i$ and to uranium dust $X^{IP}_i$ are also considered.

\subsection{Details about the Bayesian inference}

The prior distributions on the parameters defining the means of the radiological exposure variables in each cluster $c$, $c \in \mathbb{N}^*$, $\mu^{Radon}_c$, $\mu^{Gamma}_c$, $\mu^{Dusts}_c$, are informed by the data from the German cohort of uranium miners as suggested by Belloni \textit{et al.} (2020) \cite{Belloni2020}.\newline

The Bayesian inference of our BPRM model was performed with our MCMC \& PT algorithm and we applied the post-processing PAM to the MCMC outputs, as suggested to be more appropriate from our simulation study. Two Markov chains with different initial values were run. After 100 cycles of 100 iterations for the adaptive phase, the first 20,000 iterations were discarded for the burn-in phase and 40,000 additional iterations were run for each model to define the posterior sample. Trace plots of the Markov chains and the Gelman–Rubin statistics \cite{GelmanRubin1992} were used to check that there were no major convergence issues (results not shown).

\subsection{Results}

As recommended by Belloni \textit{et al.} (2020) \cite{Belloni2020}, different Markov chains were run from different initial values of the concentration parameter $\alpha$ that defines the Dirichlet process prior. Table \ref{tab:impact_postproc_donnees_reelles} gives the estimated number of non-empty clusters as well as the posterior median and 95\% credible interval of $\alpha$, for each initial position. Unlike Belloni \textit{et al.} (2020)\cite{Belloni2020}, the estimated number of non-empty clusters is very stable and does not depend on the initial position of $\alpha$ anymore, when using our MCMC \& PT algorithm. Nine non-empty clusters of uranium miners are systematically identified. Among these clusters, one group includes miners occupationally exposed to neither radon, gamma rays or uranium dust (those individuals are particularly useful to assess the instantaneous baseline hazard rate in the disease sub-model). For this cluster of so-called "non-exposed uranium miners", the excess risk of death from lung cancer is fixed to zero. \newline

\begin{table}
    \centering
    \resizebox{\columnwidth}{!}{
    \begin{tabular}{|c||c|c|c|}
    \hline
   $\alpha_{initial}$  & Estimated number of non-empty clusters & Post-median $\alpha$ & 95\% CI \\
   \hline
   0.5 & 9 & 1.13 &[0.50;2.14]\\
   1.5 & 9 & 1.17 &[0.52;2.22] \\
   2.5 & 9 & 1.05 &[0.46;2.00]\\
   3.5 & 9 & 1.07 &[0.48;2.01]\\
   4.5 & 9 & 1.10 &[0.49;2.09]\\
   5.5 & 9 & 1.09 &[0.48;2.09]\\
   6.5 & 9 & 1.05 &[0.47;2.03]\\
   7.5 & 9 & 1.07 &[0.47;2.04]\\
   \hline
    \end{tabular}
    }
    \caption{Estimated number of non-empty clusters, posterior median  and 95\% credible interval (CI) of $\alpha$ when initializing an MCMC algorithm with different values of $\alpha$, called $\alpha_{initial}$}
    \label{tab:impact_postproc_donnees_reelles}
\end{table}

\begin{figure}
    \begin{minipage}[b]{0.45\linewidth}
     \centering
     \includegraphics[width=6cm,height=6cm]{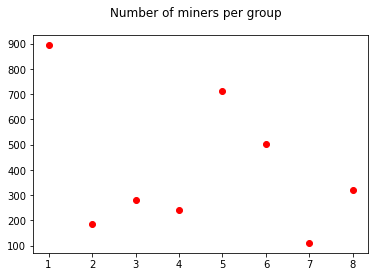} 
     \includegraphics[width=6cm,height=6cm]{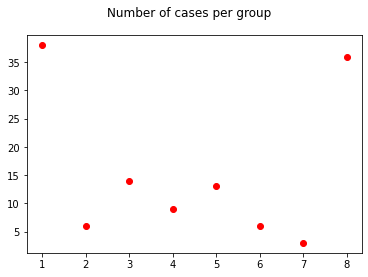} 
    \end{minipage}
  \hfill
    \begin{minipage}[b]{0.45\linewidth}
     \centering
     \includegraphics[width=6cm,height=8cm]{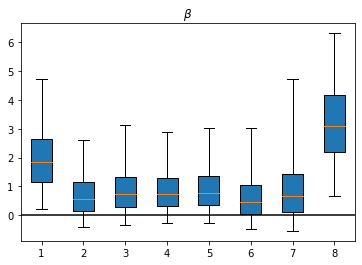}    
     \vspace{2cm}
    \end{minipage}
    \caption{Number of uranium miners per cluster (top left), number of lung cancer deaths per cluster (bottom left) and a box plot summary of the posterior distribution of the excess hazard ratio (EHR) of death from lung cancer (i.e., $\beta$) in each cluster (right), when fitting our BPRM model to the post-55 French cohort of uranium miners with our MCMC \& PT algorithm and applying the post-processing PAM. The orange line corresponds to the posterior median, the blue box is drawn between the first and third posterior quartiles and the whiskers of the boxplot show the 95\% credible interval of the posterior distribution of $\beta$ for each cluster.}
    \label{fig:nb_risk_6}
\end{figure}

On the left of Figure \ref{fig:nb_risk_6} are provided the number of uranium miners per cluster (top) as well as the number of cases of lung cancer death per cluster (bottom), except for the cluster of non-exposed uranium miners. The eight resulting clusters are denoted by 1 to 8 in the following. Clusters 1 and 5, and to a lesser extent cluster 6, include more than 60\% of uranium miners in the post-55 French cohort. However, while cluster 1 contains the highest number of cases of lung cancer death, clusters 5 and 6 include very few cases. Interestingly, cluster 8 is composed of slightly fewer individuals than clusters 1, 5 and 6 but contains many lung cancer deaths.  Cluster 7 includes the smallest number of individuals, as well as the smallest number of cases of lung cancer death. \newline 

On the right of Figure \ref{fig:nb_risk_6}, we can see a box plot summary of the posterior distribution of the excess risk $\beta$ of death from lung cancer, estimated for each of the eight clusters of exposed uranium miners. For each cluster, the blue box corresponds to the first and third posterior quartiles of $\beta$, the orange line to the posterior median and the whiskers of the box plot extend to the posterior 2.5\% and 97.5\% quantiles illustrating 95\% credible interval of $\beta$. A cluster is qualified as "high risk cluster" (or respectively "low risk") if whiskers are higher than 0 (respectively, lower than 0). Two high risk clusters of lung cancer death are here identified, namely clusters 1 and 8. The estimated excess risk of lung cancer death is equal to 3.03 with an associated 95\% credible interval equal to [0.57, 6.12] in cluster 8 and 1.68 with an associated 95\% credible interval equal to [0.04, 4.32] in cluster 1. Note that an excess risk of 1.68 means that miners belonging to this cluster have an instantaneous risk of death from lung cancer multiplied by 2.68 compared to non-exposed uranium miners. However, we can see that the above 95\% credible intervals are wide, indicating relatively high estimation uncertainty for these excess risks.   \newline 

\begin{figure}
\centering
    \includegraphics[scale = 0.75]{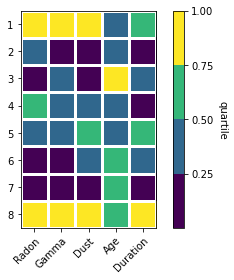}
    \caption{Comparison of the mean of the empirical distribution of each continuous exposure variable within each cluster with the quartiles of its empirical distribution over all individuals of the post-55 French cohort of uranium miners. The color purple indicates means below the 25th percentile of this distribution; the color blue indicates means between the 25th percentile and the median of this distribution; the color green indicates means between the median and the 75th percentile of this distribution and the color yellow indicates means above the 75th percentile of this distribution.}
    \label{fig:heatmap_var_quanti_6}
    \tiny
    \textsc{READING NOTE: The empirical mean of the cumulative exposure to radon in the 4th cluster is between the median and the third quartile of the empirical distribution of the cumulative exposures to radon observed for all individuals in the post-55 French cohort of uranium miners.}
\end{figure}

\begin{figure}
    \begin{minipage}[b]{0.3\linewidth}
     \centering
     \tiny
     Radon\\
     \includegraphics[width=4.4cm,height=4.4cm]{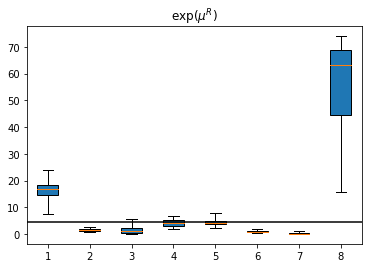}
     \includegraphics[width=4.4cm,height=4.4cm]{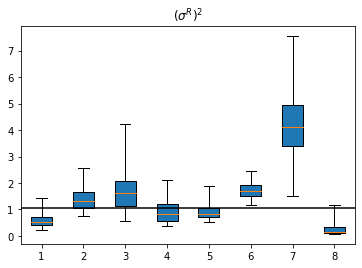} 
    \end{minipage}
  \hfill
    \begin{minipage}[b]{0.3\linewidth}
     \centering
     \tiny
     Gamma\\
     \includegraphics[width=4.4cm,height=4.4cm]{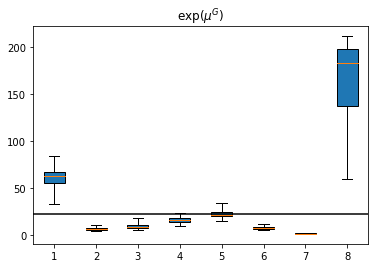}
     \includegraphics[width=4.4cm,height=4.4cm]{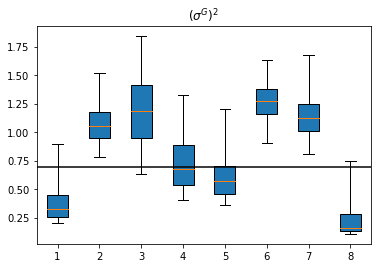}  
    \end{minipage}
    \hfill
    \begin{minipage}[b]{0.3\linewidth}
     \centering
     \tiny
     Dusts\\
     \includegraphics[width=4.4cm,height=4.4cm]{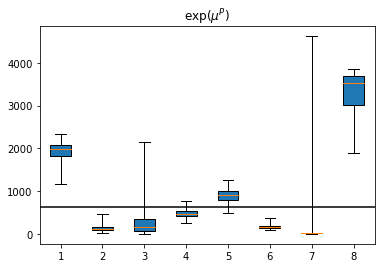}
     \includegraphics[width=4.4cm,height=4.4cm]{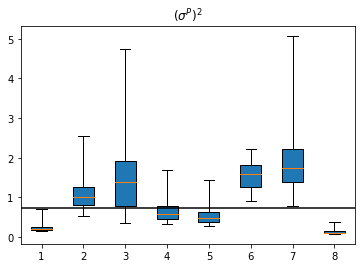}      
    \end{minipage}
    \hfill
    \begin{minipage}[b]{0.3\linewidth}
     \centering
     \tiny
    Age\\
     \includegraphics[width=4.4cm,height=4.4cm]{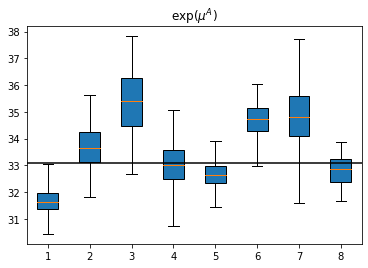}
     \includegraphics[width=4.4cm,height=4.4cm]{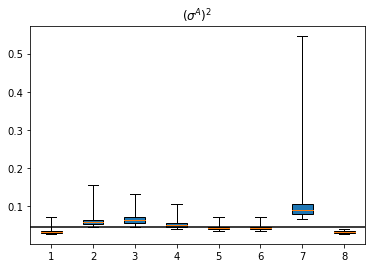}     
    \end{minipage}
    \hfill
    \begin{minipage}[b]{0.3\linewidth}
     \centering
     \tiny
     Duration\\
     \includegraphics[width=4.4cm,height=4.4cm]{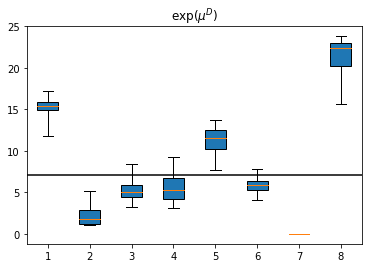}
     \includegraphics[width=4.4cm,height=4.4cm]{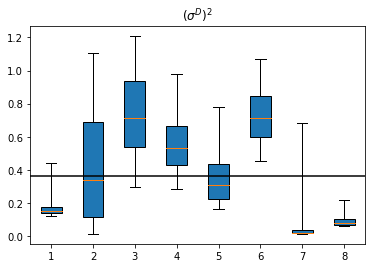}      
    \end{minipage}
    \begin{minipage}[b]{0.3\linewidth}
    \end{minipage}
    \caption{Box plot summaries of the estimated posterior distributions of the median and the variance (on log scale) of cumulative exposure to occupational radon (resp. $(exp(\mu^R)$ and $(\sigma^{R})^2$), gamma rays (resp. $(exp(\mu^G)$ and $(\sigma^{G})^2$) and uranium dust (resp. $(exp(\mu^P)$ and $(\sigma^{P})^2$), age at first exposure to one of the three radiological sources (resp. $(exp(\mu^A)$ and $(\sigma^{A})^2$) and duration of exposure (resp. $(exp(\mu^D)$ and $(\sigma^{D})^2$) within each cluster. The black line corresponds to the mean of all the cluster-specific posterior distributions.}
\label{fig:nb_param_6_quanti}
  \end{figure}

\begin{figure}
    \begin{minipage}[b]{0.32\linewidth}
     \centering
     \tiny
     Indicator radon\\
     \includegraphics[width=4.4cm,height=4cm]{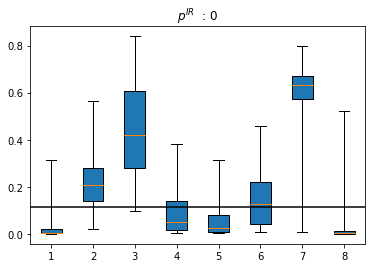}  
     \includegraphics[width=4.4cm,height=4cm]{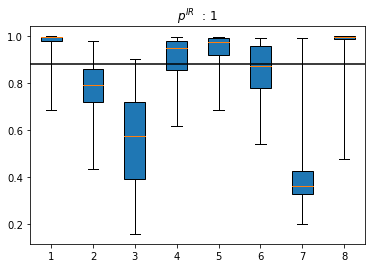}     
    \end{minipage}
    \hfill
    \begin{minipage}[b]{0.32\linewidth}
     \centering
     \tiny
     Indicator gamma
     \includegraphics[width=4.4cm,height=4cm]{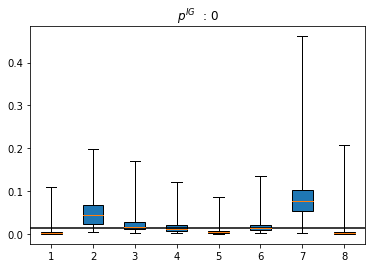}  
     \includegraphics[width=4.4cm,height=4cm]{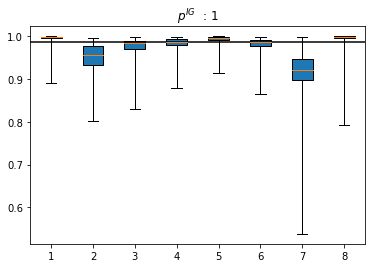}      
    \end{minipage}
    \hfill
    \begin{minipage}[b]{0.32\linewidth}
     \centering
     \tiny
     Indicator dusts
     \includegraphics[width=4.4cm,height=4cm]{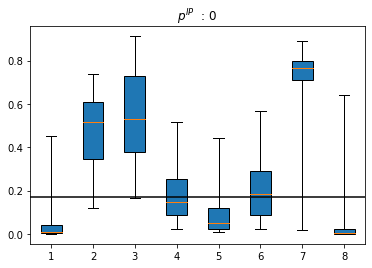}  
     \includegraphics[width=4.4cm,height=4cm]{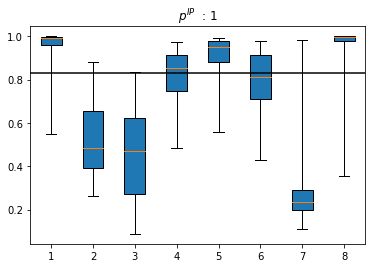}       
    \end{minipage}
    \hfill
    \begin{minipage}[b]{0.32\linewidth}
     \centering
     \tiny
     Localisation
     \includegraphics[width=4.4cm,height=4cm]{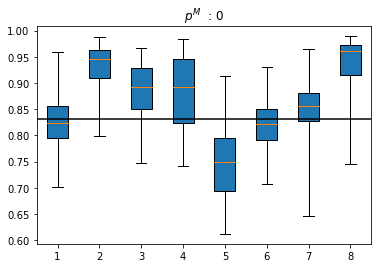}  
     \includegraphics[width=4.4cm,height=4cm]{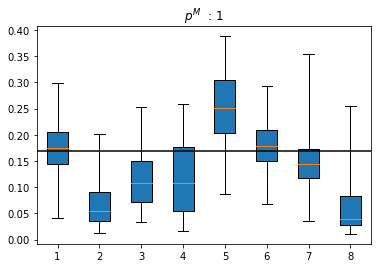}      
    \end{minipage}
    \begin{minipage}[b]{0.32\linewidth}
        \centering
        \tiny
       Job
        \includegraphics[width=4.4cm,height=1.5cm]{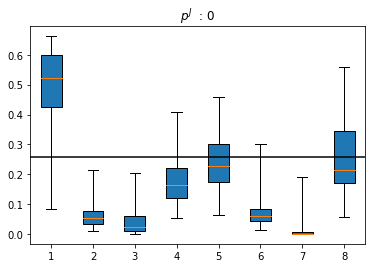}  
        \includegraphics[width=4.4cm,height=1.5cm]{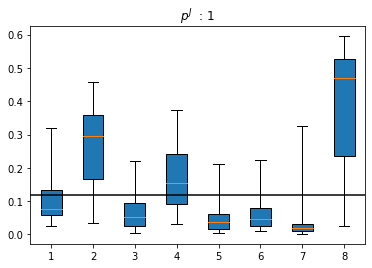}
        \includegraphics[width=4.4cm,height=1.5cm]{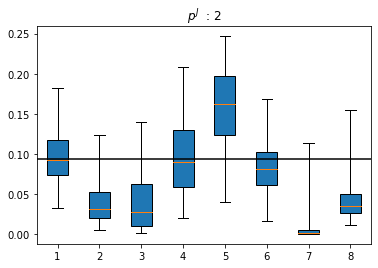}   
        \includegraphics[width=4.4cm,height=1.5cm]{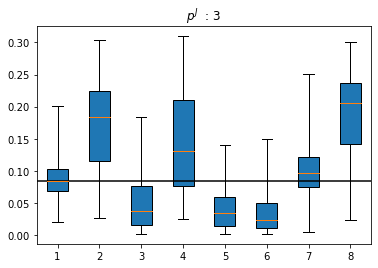} 
        \includegraphics[width=4.4cm,height=1.5cm]{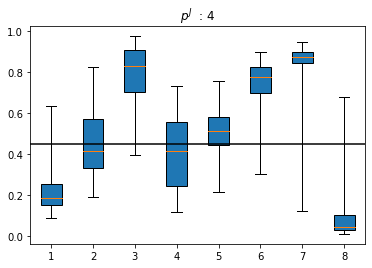}         
       \end{minipage}
       \begin{minipage}[b]{0.32\linewidth}
    \end{minipage}
    \caption{Box plot summaries of the estimated posterior distributions of the probabilities of each modality of each qualitative variable within each cluster. The black line corresponds to the mean of all the cluster-specific posterior distributions.}
\label{fig:nb_param_6_quali}
    \tiny
    \textsc{NOTE : $p^J$ : $0$ = Hewer before mechanization; $1$ = Hewer after mechanization; $2$ = other underground jobs before mechanization;
    $3$ = other underground jobs after mechanization; $4$ = surface jobs.\\
    \hspace{0.5cm} $p^M$ : $0$ = others mines; $1$ = Hérault.\\
    \hspace{0.5cm} $p^{IR}$, $p^{IG}$, $p^{IP}$ : $0$ = non exposed; $1$ = exposed. }
  \end{figure}

\begin{figure}[]
    \begin{minipage}[b]{0.5\linewidth}
     \centering
\includegraphics[width=6.5cm,height=6.5cm]{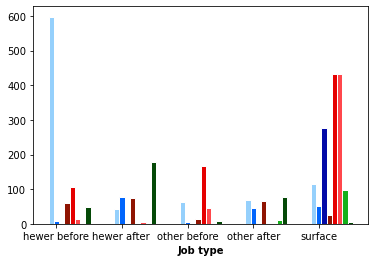}   \includegraphics[width=6.5cm,height=6.5cm]{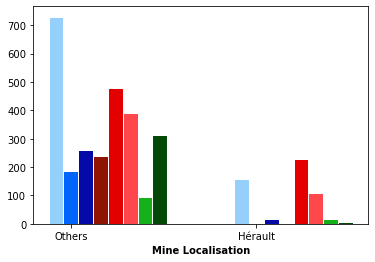}   
    \end{minipage}
    \begin{minipage}[b]{0.5\linewidth}
     \centering
\includegraphics[width=2.5cm,height=3.5cm]{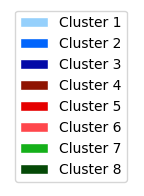}
     \vspace{1.5cm}
    \end{minipage}
    \caption{Number of uranium miners within each cluster for each modality of the qualitative exposure variables}
    \label{fig:Barplot_6}
  \end{figure}

The heat-map in Figure \ref{fig:heatmap_var_quanti_6} allows for a synthetic visual characterization of the exposure profile of the uranium miners belonging to each cluster, for continuous exposure variables. Let's consider a cluster $c$. For each continuous exposure variable $k$, it compares its empirical mean for the individuals belonging to cluster $c$ with the quartiles (i.e., 25th, 50th, 75th percentiles) of its empirical distribution observed over all individuals of the post-55 French cohort of uranium miners. Furthermore, Figures \ref{fig:nb_param_6_quanti} and \ref{fig:nb_param_6_quali} provide a more precise characterization of each cluster : they enable us to make use of the information contained in the posterior distributions, approximated by our MCMC \& PT algorithm, of the parameters that characterize the distribution of each exposure variable within each cluster. Figure \ref{fig:nb_param_6_quanti} compares box plot summaries of the approximate posterior distributions of the median ($exp(\mu^{k}_{c}$)) and the variance on logarithmic scale ($(\sigma^{k}_{c})^2$) of each continuous exposure variable within each group $c$. Boxes, orange lines, and whiskers are defined as previously. 
Figure \ref{fig:nb_param_6_quali} compares the same posterior box plot summaries within each cluster as Figure \ref{fig:nb_param_6_quanti} but for the probability of each modality of most occupied job type (i.e., hewer before and after mechanization, other underground work before and after mechanization and surface work), each modality of most occupied mine localization (i.e., Hérault and others) and the probability to be at least one year occupationally exposed to radon or gamma rays or uranium dust. In addition, Figure \ref{fig:Barplot_6} shows the number of uranium miners associated with each modality of the categorical exposure variables.\newline

Cluster 8, with the highest excess risk of lung cancer death, corresponds to the uranium miners who are the most simultaneously exposed to radon, gamma rays and uranium dust in the post-55 French cohort. In mean, their cumulative exposure to each of these three radiological sources is higher than the 75th percentile of cumulative exposures observed in this sub-cohort. Additionally, the posterior median of each radiological exposure variable is significantly higher in cluster 8. Cluster 8 is also characterized by individuals with the longest exposure durations compared to the sub-cohort as a whole. Nevertheless, age at first exposure is not a discriminating exposure variable for this cluster. Finally, even if these variables are not discriminating, cluster 8 mainly includes hewers and other underground workers after mechanization and who mainly worked in mines not located in the Hérault region.\newline 

Cluster 1, with the second highest excess risk of lung cancer death, corresponds to uranium miners who are also highly exposed to radon, gamma rays and uranium dust simultaneously, compared to the post-55 French cohort as a whole, but in a lesser extent than in cluster 8. Indeed, Figure \ref{fig:nb_param_6_quanti} shows that the posterior medians of the three radiological exposure variables are lower in cluster 1 than in cluster 8. Cluster 1 is also characterized by miners with long exposure durations compared to the other groups but shorter than cluster 8 (see Figure \ref{fig:nb_param_6_quanti}). Interestingly and contrary
to cluster 8, age at first exposure is a discriminating exposure variable
for cluster 1 : this one includes uranium miners who were young when
occupationally exposed for the first time to radon or gamma rays or
uranium dust compared to the other groups. \newline

 Clusters 2, 3, 4, 6 and 7 are composed of the uranium miners with both the lowest levels of cumulative exposure to radon, gamma rays and uranium dusts and the shortest durations of exposure observed in the post-55 French cohort. Age at first exposure is not a discriminating variable . However, we can notice that it is rather high in these 5 clusters (except for cluster 4). Finally, we can notice that clusters 3, 6, and 7 are predominantly composed of miners working at the surface, which explains in part their lower exposure to ionizing radiation. Due to the fact that cluster 7 contains few miners (i.e., about 100 individuals), we can notice that all the parameters associated to this cluster are not precisely estimated in comparison with other clusters. \newline  
 
Cluster 5 is composed of individuals who are less exposed to radon and gamma radiation but rather highly exposed to uranium dusts compared to the cohort as a whole. Exposure duration is a discriminating variable for this cluster that is characterized by miners with long exposure durations compared to the other groups.\newline

Table \ref{tab:résumé_clusters} is derived from Figures \ref{fig:nb_param_6_quanti} and \ref{fig:nb_param_6_quali} : it summarizes the main characteristics (exposure profiles) of the eight identified clusters with the symbols -{}- , -, +, ++ for respectively ‘much lower’, ‘lower’, “higher”, “much higher” than the mean of the cluster-specific posterior distributions of continuous exposure variables and the most occupied job type. We set the symbol ++ (resp. -{}-) when the $95\%$ credible interval associated to one cluster-specific posterior distribution is strictly lower (resp. higher) than the mean of all the cluster-specific posterior distributions.  \newline

\begin{table}
\resizebox{\textwidth}{!}{
    \begin{tabular}{|c||c|c|c|c|c|c|}
        \hline
        Cluster & Radon & Gamma &  Dusts & Age & Duration & Job \\
        \hline
        \hline
        $1$ & ++ & ++ & ++ & -{}- & ++ & Hewer before\\
        \hline
        $2$ & -{}- & -{}- & -{}- & + & -{}- & Underground after and  Hewer after \\
        \hline
        $3$ & - &  -{}- & - & +  & - & Surface \\
        \hline
        $4$ & - & -{}- & - & - & - & Underground after \\
        \hline
        $5$ & - & - & + & - & ++ & Underground before \\
        \hline
        $6$ & -{}- & -{}- & -{}- & + & - & Surface \\
        \hline
        $7$ & -{}- & -{}- & - & + & -{}- & Surface \\
        \hline
        $8$ & ++ & ++ & ++ & - & ++ &  Underground after and Hewer after\\
        \hline
    \end{tabular}
    }
    \caption{Characterisation of the exposure profiles associated to each non-empty cluster of exposed uranium miners when fitting a BRPM model to the post-55 French cohort of uranium miners. The symbols -{}- , -, +, ++ means respectively ‘much lower’, ‘lower’, “higher”, “much higher” than the mean of all the cluster-specific posterior distributions.}
   \tiny
    \textsc{Note : "Before" stands for "before mechanization" and "after" stands for "after mechanization"}
    \label{tab:résumé_clusters}
\end{table}

\section{Discussion}

In this paper, we propose a MCMC algorithm coupled with a parallel tempering algorithm (MCMC \& PT algorithm) to fit a BPRM model based on censored survival health outcomes. MCMC algorithms are a popular technique in Bayesian inference. However, there are known to be slow, especially for high-dimensional models based on a huge number of latent variables, like BPRM models. In order to accelerate the inference, we propose a parallelisation of the update of the latent variables of the algorithm, namely the class labels and the auxiliary variables defined in the slice sampler \cite{Walker2007, Liverani2015}. We provide a code in JAX leading in our case to a decrease of the CPU/GPU time of 94\%.  A simulation study is performed to compare three methods of post-processing of the MCMC outputs in terms of estimated number of non-empty clusters, proportion of misclassified individuals and relative bias on the risk parameters. Our methods are applied to the estimation of the excess risk of lung cancer death associated with multiple occupational exposures to ionizing radiation in the post-55 sub-cohort of French uranium miners.\newline

Regarding the above case study, our MCMC \& PT algorithm allows to estimate a stable number of non-empty clusters, compared to Belloni et al. (2020)\cite{Belloni2020}. Our algorithm allows to identify two high risk clusters of lung cancer death in the post-55 sub-cohort of French uranium miners, with similar exposure profiles as in Belloni et al. (2020)\cite{Belloni2020}. The first group (cluster 8) corresponds to the uranium miners the most simultaneously and highly exposed to radon, gamma rays and uranium dust and with the longest durations of exposure (more than 20 years) compared to the post-55 sub-cohort of uranium miners as a whole (mainly after mechanization and not in the mine located at Herault). The second group (cluster 1) corresponds to the uranium miners who were very young (less than 32 years) when occupationally exposed for the first time to radon or gamma rays or uranium dust and who were highly exposed to these three sources of radiological exposure for more than 15 years (mainly hewer before mechanization). Nevertheless, we can see that the excess hazard ratio (EHR) of lung cancer death estimated for each of these two high risk clusters are higher and marked by a larger posterior uncertainty in the present work than in Belloni et al. (2020)\cite{Belloni2020}.  In the first group (cluster 8), the posterior median of the EHR is 3.10 with a 95\%CI= [0.68; 6.31] compared to 1.4 with a 95\%CI = [0.60; 2.60] in Belloni et al. (2020)\cite{Belloni2020}. In the second group (cluster 1), the posterior median of the EHR is 1.86 with a 95\%CI= [0.23; 4.72] compared to 1.2 with a 95\%CI = [0.17; 2.80] in Belloni et al. (2020)\cite{Belloni2020}. These differences can be explained firstly by the fact that the study populations are not exactly the same. Indeed, the data used in this paper are those from the latest update of the French cohort of uranium miners, which includes an additional seven years of follow-up of miners compared to Belloni et al. (2020)\cite{Belloni2020}. Furthermore, our MCMC \& PT algorithm allows to perform a full Bayesian inference of the BPRM model. This implies to explicitly account for the uncertainty on the number of clusters during the inference process - and therefore in the excess risk estimates - whereas this number is fixed in Belloni et al. (2020)\cite{Belloni2020}. Accounting for this additional and important source of uncertainty may explain why the posterior uncertainty on the cluster-specific EHRs of lung cancer death is estimated to be higher in the present work. Finally, note that, for the same reasons as the ones explained in Belloni et al. (2020)\cite{Belloni2020}, it was not possible to account for the tobacco consumption of miners in this study which does not, however, call into question the obtained results. \newline 

In light of the results obtained in our simulation study, we would recommend the use of the partitioning around medoids (PAM) algorithm to post-process the output of any MCMC algorithm, if one aim is to find the true number of clusters when fitting a BPRM model. Despite the fact that this method is \textit{ad hoc} and not designed for DPM models, it allows both to smooth the uncertainty on the clustering at the end of the MCMC algorithms and to delete the "small" clusters induced by the use of the Dirichlet process prior. Minimising the variation of information (VI) is a judicious choice of post-treatment if the objective is not to recover the true partitioning but to identify stable estimated clusters, even if it means grouping clusters together when there is uncertainty. In our simulation study, this post-processing method tends to gather clusters that are closed to each other. However, the partitioning is stable from one scenario to another.\newline

BPRM models do not tend to classify individuals that are not at risk in clusters at risks. However, it can happen that an important proportion of individuals at risk are classified in groups that are not at risk. In order to decrease this proportion of misclassification, we recommend to carry on the inference with a MCMC algorithm coupled with a parallel tempering algorithm. If the different chains required by the parallel tempering algorithm are run in parallel, the temporal complexity of the inference algorithm is not much higher than a classic MCMC algorithm. However, the spatial complexity of the algorithm increased. It is not an issue for our case study. \newline

In this work, the health risk can be informed by survival outcome data. This kind of data may be more or less censored, leading to a more or less significant loss of information. In the case where the information provided by the health outcome is much lower than the information provided by the exposure covariates, the clustering can only be informed by the exposure covariates. This can lead to a high proportion of individuals classified in clusters with different health risks. In our simulation study, for instance, we believe that the proportion of misclassified individuals would be lower with more informative outcome data, like categorical data (and the use of a multinomial regression) or continuous data (and the use of a Gaussian linear regression).\newline

BPRM models are used to classify individuals into groups with similar exposure profiles and a similar health risk of interest. However, it is important to note that the class labels  are latent variables that have been artificially defined for the needs of the model. The actual existence of groups in the French cohort of uranium miners is highly unlikely. This undoubtedly explains why the clustering is more stable when using data simulated according to a finite mixture model compared to the observational data of the French cohort.\newline

It would be interesting to quantify the influence of each exposure variable in the excess risk estimate obtained for each cluster. However, this is not trivial when using a BPRM model.
And more generally, the problem of measuring the influence of each exposure when estimating a health risk in a context of multiple exposures is an open research topic called the attributable share of each exposure. Liverani \textit{et al.} (2015) \cite{Liverani2015} propose to assess if the probability distribution of each exposure variable differs from one cluster to another, or not. They use this result to select the variables to be included in the exposure profiles. An extension of this work could be proposed in order to measure the share of each variable in the construction of each cluster. \newline 

Competitive risks are ignored in the set of survival models proposed in this work. However, in the specific survival context considered in this work, it is not relevant since the estimated instantaneous risk function (and therefore the estimated instantaneous risk ratio) is not biased by the presence of competing risks (unlike a cumulative incidence function or a survival function) \cite{Andersen2012}.\newline 

Our BPRM model does not take into account the temporal dimension of exposure to radon, gamma rays and uranium dusts. We only consider the cumulative exposure for each covariate, over the whole of the miner's exposure period.
This modelling choice led us to 
set the time of left truncation of the survival outcome variable to the time of last exposure. This is because using survival models requires covariates to be fixed in time when individuals enter the cohort.  One new avenue for methodological research in BPRM models could be to extend them to deal with time-dependent exposure covariates. \newline

BPRM models can be used in the context of assessing a health risk induced by some exposome. However, Guillien \textit{et al.} ($2021$) \cite{Guillien2021} emphasise that assessing the exposures of an exposome is an error-prone process. BPRM models could be extended to account for exposure measurement error \cite{Hoffmann2017, Belloni2020b}. However, this would imply a major increase of the number of latent variables and would significantly increase the time complexity of the MCMC algorithms. For this perspective, our work is crucial since it already significantly reduces the temporal complexity of the MCMC algorithm when not accounting for measurement error.\newline

In this work, we focus on a case study on occupational exposure to several sources of ionizing radiations of French uranium miners, considering only a small number (i.e., 7) of exposure covariates. It would be interesting to challenge the performances of our parallelized MCMC \& PT algorithm in case studies where the number of exposure covariates would be clearly higher (like, for instance, environmental and/or genetic risk factors).




\newpage

\section*{Supplementary material}
\begin{figure}[H]
    \centering
    \includegraphics[width=0.8\linewidth]{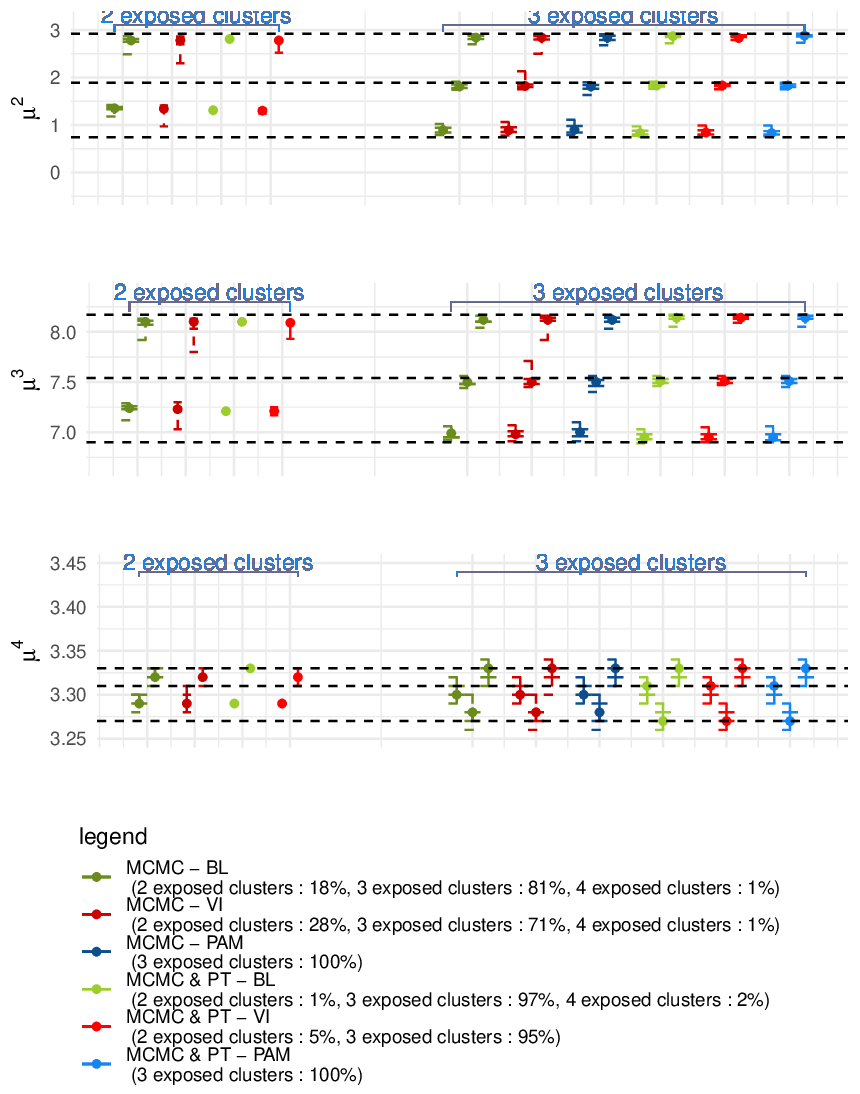}
    \caption{Means (points), 75\% (solid lines) and 95\% (dashed lines) uncertainty intervals of the 100 posterior means estimated for $\mu_2$ (at the top), $\mu_3$ (at the center) and $\mu_4$ (at the bottom) in each estimated exposed cluster, for datasets simulated under the scenario S1 and given the estimated number of exposed clusters. The horizontal dashed lines correspond to the true values of the parameters for the three true exposed clusters.}
    \label{fig:mean_param_annexe}
\end{figure}

\begin{figure}[H]
    \centering
    \includegraphics[width=1\linewidth]{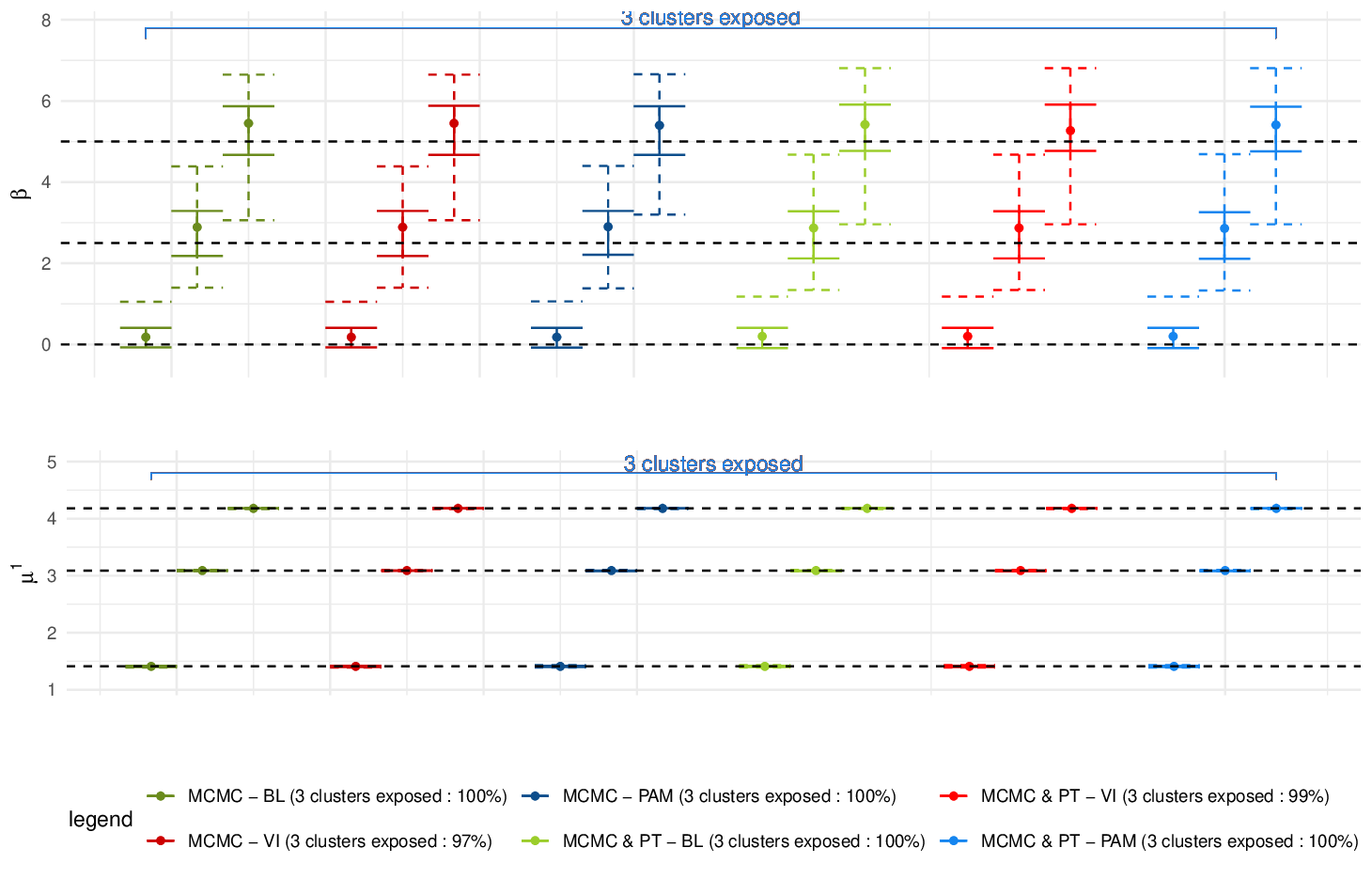}
    \caption{Means (points), 75\% (solid lines) and 95\% (dashed lines) credible intervals of the 100 posterior means estimated for $\beta$ (at the top) and $\mu^1$ (at the bottom) in each estimated exposed cluster, for datasets simulated under the scenario S2 and given the estimated number of exposed clusters. The horizontal dashed lines correspond to the true values of the parameters for the three true exposed clusters.}
    \label{fig:mean_param_S2}
\end{figure}

\begin{figure}[H]
    \centering
    \includegraphics[width=1\linewidth]{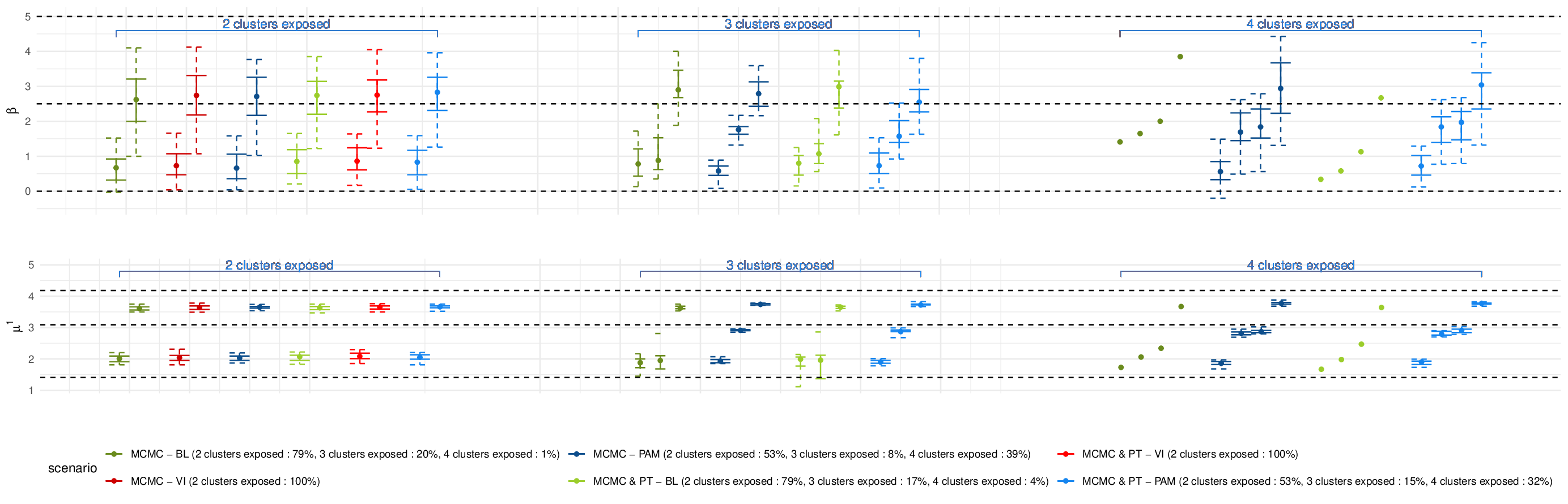}
    \caption{Means (points), 75\% (solid lines) and 95\% (dashed lines) credible intervals of the 100 posterior means estimated for $\beta$ (at the top) and $\mu^1$ (at the bottom) in each estimated exposed cluster, for datasets simulated under the scenario S3 and given the estimated number of exposed clusters. The horizontal dashed lines correspond to the true values of the parameters for the three true exposed clusters.}
    \label{fig:mean_param_S3}
\end{figure}

\begin{figure}[H]
    \centering
    \includegraphics[width=1\linewidth]{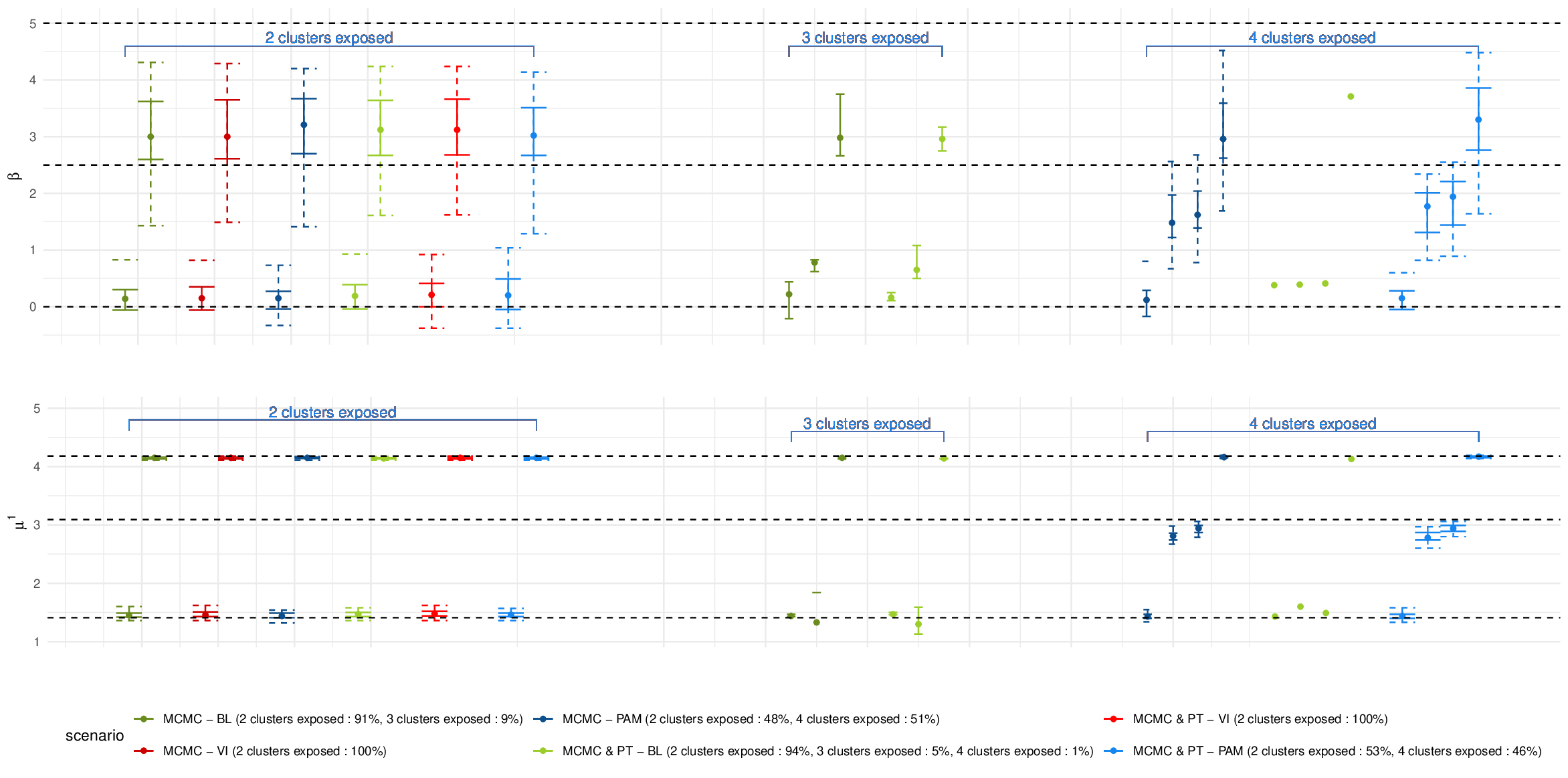}
    \caption{Means (points), 75\% (solid lines) and 95\% (dashed lines) credible intervals of the 100 posterior means estimated for $\beta$ (at the top) and $\mu^1$ (at the bottom) in each estimated exposed cluster, for datasets simulated under the scenario S4 and given the estimated number of exposed clusters. The horizontal dashed lines correspond to the true values of the parameters for the three true exposed clusters.}
    \label{fig:mean_param_S4}
\end{figure}

\begin{figure}[H]
    \centering
    \includegraphics[width=0.7\linewidth]{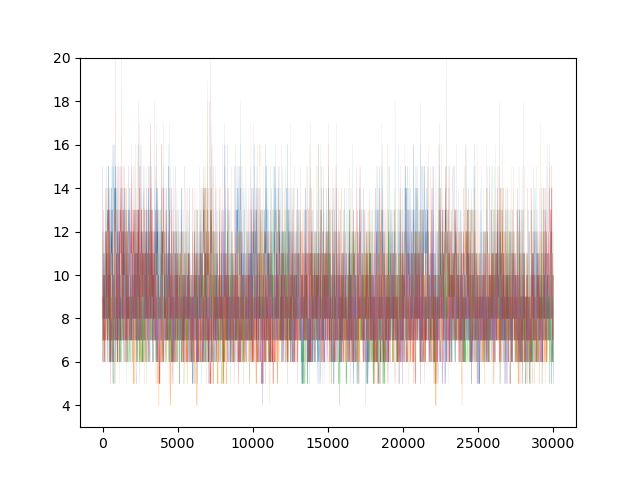}
    \caption{Traces of the Markov chains of the number of non-empty clusters obtained from the classical MCMC algorithm and applied to the post-55 French cohort of uranium miners}
    \label{fig:enter-label}
\end{figure}

\begin{figure}[H]
    \centering
    \includegraphics[width=0.7\linewidth]{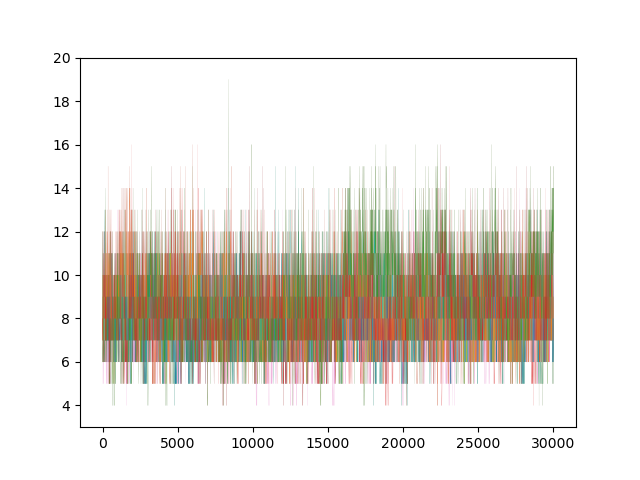}
    \caption{
    Traces of the Markov chains of the number of non-empty clusters obtained from our MCMC algorithm coupled with a parallel tempering algorithm and applied to the post-55 French cohort of uranium miners}
    \label{fig:enter-label2}
\end{figure}


\begin{thebibliography}{99}

\bibitem{HwangLee2005}
Hwang~SY and Lee~JH. Comparison of cardiovascular risk profile clusters among industrial workers. \textit{Taehan Kanho Hakhoe Chi.} 2005; 35(8): 1500-7. DOI: 10.4040/jkan.2005.35.8.1500.

\bibitem{Ahlqvist2018}
Ahlqvist~E, Storm~P, Käräjämäki~, Martinell~M, Dorkhan~M, Carlsson~A, Vikman~P, Prasad~RB, Aly~DM, Almgren~P, Wessman~Y, Shaat~N, Spégel~P, Mulder~H, Lindholm~E, Melander~O, Hansson~O, Malmqvist~U, Lernmark~Å, Lahti~K, Forsén~T, Tuomi~T, Rosengren~AH and Groop~L. Novel subgroups of adult-onset diabetes and their association with outcomes: a data-driven cluster analysis of six variables. \textit{Lancet Diabetes Endocrinol.} 2018; 6(5): 361-369. DOI: 10.1016/S2213-8587(18)30051-2

\bibitem{Harmouche-Karaki2019}
Harmouche-Karaki~M, Mahfouz~Y, Salameh~P, Matta~J, Helou~K and Narbonne~JF. Patterns of PCBs and OCPs exposure in a sample of Lebanese adults: The role of diet and physical activity. \textit{Environmental Research} 2019; 179: 108789. DOI: 10.1016/j.envres.2019.108789

\bibitem{Guillien2021b}
Guillien~A, Lepeule~J, Seyve~E, Le Moual~N, Pin~I, Degano~B, Garcia-Aymerich~J, Pépin~JL, Pison~C, Dumas~O, Varraso~R and Siroux~V. Profile of exposures and lung function in adults with asthma: An exposome approach in the EGEA study. \textit{Environ. Res.} 2021; 196: 110422. DOI: 10.1016/j.envres.2020.110422


\bibitem{Guillien2022}
Guillien~A, Bédard~A, Dumas~O, Allegre~J, Arnault~N, Bochaton~A, Druesne-Pecollo~N,  Dumay~D, Fezeu~LK, Hercberg~S, Le Moual~N, Pilkington~H, Rican~S, Sit~G, Szabo de Edelenyi~F, Touvier~M, Galan~P, Feuillet~T, Varraso~R and Siroux~V. Exposome Profiles and Asthma among French Adults. \textit{American Journal of Respiratory and Critical Care Medicine} 2022; 10(206): 1208-1301. DOI: 10.1164/rccm.202205-0865OC

\bibitem{Wade2023}
Wade~S. Bayesian cluster analysis. \textit{Philos. Trans. A Math. Phys. Eng. Sci.} 2023; 381(2247): 20220149. DOI: 10.1098/rsta.2022.0149

\bibitem{Rage2015}
Rage~E, Caër-Lorho~S, Drubay~D, Ancelet~S, Laroche~P and Laurier~D. Mortality analyses in the updated French cohort of uranium miners (1946–2007). \textit{Int. Arch. Occup. Environ. Health} 2015; 88: 717–730. DOI: 10.1007/s00420-014-0998-6

\bibitem{Belloni2020}
Belloni~M, Laurent~O,  Guihenneuc~C and Ancelet~S. Bayesian Profile Regression to Deal With Multiple Highly Correlated Exposures and a Censored Survival Outcome. First Application in ionising Radiation Epidemiology, \textit{Front. Public Health} 2020; 8: 557006. DOI: 10.3389/fpubh.2020.557006

\bibitem{Majcherek2021}
Majcherek~D, Weresa~M.A. and Ciecierski~C. A Cluster Analysis of Risk Factors for Cancer across EU Countries: Health Policy Recommendations for Prevention. \textit{Int. J. Environ. Res. Public Health} 2021; 18: 8142. DOI: 10.3390/ijerph18158142

\bibitem{Ma2023}
Ma~T, Tu~K, Ou~Q, Fang~Y and Zhang~C. Comparing the Associations of Dietary Patterns Identified through Principal Component Analysis and Cluster Analysis with Colorectal Cancer Risk: A Large Case-Control Study in China, \textit{Nutrients} 2023; 16(1): 147. DOI: 10.3390/nu16010147

\bibitem{Chuang2024}
Chuang~YC, Tsai~HH, Lin~MC, Wu~CC, Lin~YC and Wang~TN. Cluster analysis of phenotypes, job exposure, and inflammatory patterns in elderly and nonelderly asthma patients. \textit{Allergology International} 2024; 73(2): 214-223, ISSN 1323-8930, DOI: 10.1016/j.alit.2024.01.001.

\bibitem{Molitor2010}
Molito~J, Papathomas~M, Jerrett~M and Richardson~S. Bayesian profile regression with an application to the National survey of children's health. \textit{Biostatistics} 20210; 11(3): 484-498, DOI: 10.1093/biostatistics/kxq013

\bibitem{Molitor2011}
Molitor~J, Su~JG, Molitor~NT, G'{o}mez Rubio~V, Richardson~S, Hastie~D, Morello-Frosch~R and Jerrett~M. Identifying Vulnerable Populations through an Examination of the Association Between Multipollutant Profiles and Poverty. \textit{Environ. Sci. Technol.} 2011; 45: 7754–7760. DOI: 10.1021/es104017x

\bibitem{Hastie2013}
Hastie~DI, Liverani~S, Azizi~L, Richardson~S and Stücker~I. A semi-parametric approach to estimate risk functions associated with multi-dimensional exposure profiles: application to smoking and lung cancer. \textit{BMC Medical Research Methodology} 2013; 13: 129. DOI: 10.1186/1471-2288-13-129

\bibitem{Pirani2015}
Pirani~M, Best~N, Blangiardo~M, Liverani~S, Atkinson~RW and Fuller~GW. Analysing the health effects of simultaneous exposure to physical and chemical properties of airborne particles. \textit{Environment International} 2015; 79: 56-64. DOI: 10.1016/j.envint.2015.02.010

\bibitem{Coker2016}
Coker~E, Liverani~S, Ghosh~JK, Jerrett~M, Beckerman~B, Li~A, Ritz~B and Molitor~J. Multi-pollutant exposure profiles associated with term low birth weight in Los Angeles County. \textit{Environment International} 2016;  91: 1-13. DOI: 10.1016/j.envint.2016.02.011

\bibitem{Coker2018}
Coker~E, Liverani~S, Su~JG and Molitor~J. Multi-pollutant Modeling Through Examination of Susceptible Subpopulations Using Profile Regression. \textit{Current Environmental Health Reports} 2018; 5: 59–69. DOI: 10.1007/s40572-018-0177-0

\bibitem{Lavigne2020}
Lavigne~A, Freni-Sterrantino~A, Fecht~D, Liverani~S, Blangiardo~M, de Hoogh~K, Molitor~J and Hansell~A. A spatial joint analysis of metal constituents of ambient particulate matter and mortality in England. \textit{Environmental Epidemiology} 2020; 4(4): e098. DOI: 10.1097/EE9.0000000000000098


\bibitem{Molitor2014}
Molitor~J, Brown~IJ, Chan~Q, Papathomas~M, Liverani~S, Molitor~N, Richardson~S, Van Horn~L, Daviglus~ML, Dyer~A, Stamler~J and Elliott~P. Blood Pressure Differences Associated With Optimal Macronutrient Intake Trial for Heart Health (OMNIHEART)–Like Diet Compared With a Typical American Diet. \textit{Hypertension} 2014; 4(6): 1198-1204. DOI: 10.1161/HYPERTENSIONAHA.114.03799


\bibitem{Rouanet2024}
Rouanet~A, Johnson~R, Strauss~M, Richardson~S, Tom~BD, White~SR and Kirk~PDW. Bayesian profile regression for clustering analysis involving a longitudinal response and explanatory variables. \textit{Journal of the Royal Statistical Society Series C: Applied Statistics} 2024; 4(73): 314–339. DOI: 10.1093/jrsssc/qlad097

\bibitem{Liverani2021}
Liverani~S, Leigh~L, Hudson~IL, Byles~JE. Clustering method for censored and collinear survival data. \textit{Computational Statistics} 2021; 36:35–60. DOI: 10.1007/s00180-020-01000-3

\bibitem{Fendler2024}
Fendler~J, Guihenneuc~C and Ancelet~S. Bayesian identification and estimation of radon-related increased hazard rates of cancer death in the updated French cohort of uranium miners (1946–2014). \textit{Int. Arch. Occup. Environ. Health} 2024; DOI: 10.1007/s00420-024-02098-4


\bibitem{Liverani2015}
Liverani~S, Hastie~DI, Azizi~L, Papathomas~M and Richardson~S. PReMiuM: An R Package for Profile Regression Mixture Models Using Dirichlet Processes. \textit{J. Stat. Softw.} 2015; 7(64):1-30. DOI: 10.18637/jss.v064.i07

\bibitem{Giampiccolo2024}
Giampiccolo~C, Amado~A, Coudon~T, Praud~D, Grassot~L, Faure~E, Couvidat~F, Severi~G, Mancini~FR, Fervers~B and Roy~P. Multi-pollutant exposure profiles associated with breast cancer risk: A Bayesian profile regression analysis in the French E3N cohort. \textit{Environment International} 2024; 190:108943. DOI: 10.1016/j.envint.2024.108943

\bibitem{Geyer1992}
Geyer~CJ. Practical Markov Chain Monte Carlo. \textit{Statistical Science} 1992; 7(4): 473-483

\bibitem{Napier2019}
Napier~G, Lee~D, Robertson~C and Lawson~A. A Bayesian space–time model for clustering areal units based on their disease trends. \textit{Biostatistics} 2019; 20(4): 681-697. DOI: 10.1093/biostatistics/kxy024

\bibitem{Wade2018}
Wade~S and Ghahramani~Z. Bayesian Cluster Analysis: Point Estimation and Credible Balls (with Discussion). \textit{Bayesian Analysis} 2018; 13(2): 559-629, DOI: 10.1214/17-BA1073


\bibitem{Pitman2002}
Pitman~J. \textit{Combinatorial stochastic processes, Technical report} 2002; Dept. Statistics UC Berkeley.


\bibitem{Hoffmann2017}
Hoffmann~S, Rage~E, Laurier~D, Laroche~P, Guihenneuc~C and Ancelet~S. Accounting for Berkson and Classical Measurement Error in Radon Exposure Using a Bayesian Structural Approach in the Analysis of Lung Cancer Mortality in the French Cohort of Uranium Miners. \textit{Radiat. Res.} 2017; 187(2): 196-209. DOI: 10.1667/RR14467.1. 

\bibitem{Ascolani2023}
Ascolani~, Lijoi~A, Rebaudo~G and Zanella~G. Clustering consistency with Dirichlet process mixtures. \textit{Biometrika} 2023; 110(2): 551–558. DOI: 10.1093/biomet/asac051

\bibitem{GuhaHoNguyen2021}
Guha~A, Ho~N and Nguyen~X. On posterior contraction of parameters and interpretability in Bayesian mixture modeling. \textit{Bernoulli} 2021; 27(4): 2159-2188. DOI: 10.3150/20-BEJ1275

\bibitem{Vose1996}
Vose~D. \textit{Quantitative Risk Analysis: A Guide to Monte Carlo Simulation Modelling}. John Wiley \& Sons 1996

\bibitem{Walker2007}
Walker~S. Sampling the Dirichlet Mixture Model with Slices. \textit{Communications in Statistics - Simulation and Computation} 2007; 36(1): 45-54, DOI: 10.1080/03610910601096262

\bibitem{Stephens2000}
Stephens M. Dealing With Label Switching in Mixture Models. \textit{Journal of the Royal Statistical Society Series B: Statistical Methodology} 2000; 62(4): 795–809. DOI: 10.1111/1467-9868.00265

\bibitem{HastieLiveraniRichardson2015}
Hastie~DI, Liverani~S and Richardson~S. Sampling from Dirichlet process mixture models with unknown concentration parameter: mixing issues in large data implementations. \textit{Stat. Comput.} 2014; 25:1023-1037. DOI: 10.1007/s11222-014-9471-3


\bibitem{Sambridge2014}
Sambridge~M. A Parallel Tempering algorithm for probabilistic sampling and multimodal optimization. \textit{Geophysical Journal International} 2014; 196(1): 357-374, DOI: 10.1093/gji/ggt342



\bibitem{EarlDeem2005}
Earl~DJ and Deem~MW. Parallel tempering: Theory, applications, and new perspectives. \textit{Phys. Chem. Chem. Phys.} 2005; 7(23): 3910-3916. DOI: 10.1039/B509983H


\bibitem{Laloy2016}
Laloy~E, Linde~N, Diederik~J and Mariethoz~G. Merging parallel tempering with sequential geostatistical resampling for improved posterior exploration of high-dimensional subsurface categorical fields. \textit{Advances in Water Resources} 2016; 90: 57-69. DOI: 10.1016/j.advwatres.2016.02.008


\bibitem{jax2018github}
Bradbury~J, Frostig~R, Hawkins~P, Johnson~MJ, Leary~C, Maclaurin~D, Necula~G, Paszke~Ã, Vander{P}las~J, Wanderman-{M}ilne~S, Zhang~Q,{JAX}: composable transformations of {P}ython+{N}um{P}y programs 0.3.13 2018; http://github.com/jax-ml/jax

\bibitem{FritschIckstadt2009}
Fritsch~A and Ickstadt~K. Improved criteria for clustering based on the posterior similarity matrix. \textit{Bayesian Anal.} 2009; 4(2): 367-391, DOI: 10.1214/09-BA414

\bibitem{KaufmanRousseeuw1987}
Kaufman~L and Rousseeuw~PJ. Clustering by means of medoids. \textit{Data Analysis Based on the L1-Norm and Related Methods} 1987; 405-416

\bibitem{Meila2007}
Meilă~M. Comparing clusterings—an information based distance. \textit{Journal of Multivariate Analysis} 2007; 98(5): 873-895. DOI: 10.1016/j.jmva.2006.11.013

\bibitem{mcclust.ext}
Wade~S and Ghahramani~Z. \textit{mcclust.ext 1.0.tar} 2015; R package

\bibitem{Vacquier2011}
Vacquier~B,  Rage~E, Leuraud~K, Caër-Lorho~S, Houot~J, Acker~A and Laurier~D. The Influence of Multiple Types of Occupational Exposure to Radon, Gamma Rays and Long-Lived Radionuclides on Mortality Risk in the French "post-55" Sub-cohort of Uranium Miners: 1956–1999. \textit{Radiat. Res.} 2011; 176(6): 796-806. DOI: 10.1667/rr2558.1

\bibitem{UNSCEAR2006}
UNSCEAR \textit{Effects of ionizing radiation, UNSCEAR (United Nations Scientific Committee on the Effects of Atomic Radiation) 2006 Report to the General Assembly with scientific Annexes, Volume 1 Annex A}


\bibitem{GelmanRubin1992}
Gelman, A and Rubin, DB. Inference from Iterative Simulation Using Multiple Sequences. \textit{Statistical Science} 1992; 7(4): 457–472

\bibitem{Andersen2012}
Andersen~PK, Geskus~RB, de Witte~T and Putter~H. Competing risks in epidemiology: possibilities and pitfalls. \textit{American Journal of Epidemiology} 2012; 41: 861-870. DOI: 10.1093/ije/dyr213

\bibitem{Guillien2021}
Guillien~A, Cadiou~S, Slama~R and Siroux~V. The Exposome Approach to Decipher the Role of Multiple Environmental and Lifestyle Determinants in Asthma. \textit{Int. J. Environ. Res. Public Health} 2021; 18(3). DOI: 10.3390/ijerph18031138

\bibitem{Belloni2020b}
Belloni~M, Guihenneuc~C, Rage~E and Ancelet~S. A Bayesian hierarchical approach to account for left-censored and missing radiation doses prone to classical measurement error when analyzing lung cancer mortality due to $\gamma$-ray exposure in the French cohort of uranium miners. \textit{Radiat. Environ. Biophys.} 2020; 59(3): 423-437. DOI: 10.1007/s00411-020-00859-6









\end{thebibliography}
\end{document}